\documentclass[aps,prd,two column]{revtex4-2}

\usepackage{amsmath}
\usepackage[normalem]{ulem}
\usepackage{color}
\usepackage{graphicx}
\usepackage{epsfig}
\usepackage{subfig}
\usepackage{bm}
\usepackage{hyperref}
\usepackage{natbib}

\def\be {\begin{equation}}
\def\ee {\end{equation}}
\def\nn {\nonumber}
\def\bea {\begin{eqnarray}}
\def\eea {\end{eqnarray}}

\newcommand{\ep}{\epsilon}
\newcommand{\om}{\omega} 

\newcommand{\vk}{\vec k}

\newcommand{\del}{\partial}

\begin{document}

\title{Thermal conductivity of evolving quark-gluon plasma in the presence of a time-varying magnetic field}

\author{Kamaljeet Singh}
\email{kspaink84@gmail.com}
\author{Jayanta Dey}
\email{jayantad@iitbhilai.ac.in}
\author{Raghunath Sahoo\footnote{Corresponding Author: Raghunath.Sahoo@cern.ch}}
\email{Corresponding Author: Raghunath.Sahoo@cern.ch}
\affiliation{ Department of Physics, Indian Institute of Technology Indore, Simrol, Indore 453552, India}

\begin{abstract}
The effect of the temperature evolution of QGP on its thermal conductivity and elliptic flow is investigated here in the presence of a time-varying magnetic field. Thermal conductivity plays a vital role in the cooling rate of the medium or its temperature evolution. The magnetic field produced during the early stages of (non-central) heavy-ion collisions decays with time, where electrical conductivity plays a significant role. As the medium expands, the electrical and thermal properties change, reflecting the effect in various observables. In this study, we have calculated the thermal conductivity of the QGP medium, incorporating the effects of temperature and magnetic field evolution. We discovered that conductivity significantly depends on the cooling rate, and its value increases due to temperature evolution. Furthermore, the influence of these evolutions on the elliptic flow coefficient is measured, and elliptic flow decreases due to the evolution. We also extend our study for the case of Gubser flow, where, along with the longitudinal Bjorken expansion, the radially transverse expansion is also present. 
\end{abstract}

\maketitle
	\section{Introduction}

 The experimental facilities Large Hadron Collider (LHC) and Relativistic Heavy Ion Collider (RHIC) have successfully produced the quark-gluon plasma (QGP) during the heavy ion collision experiments~\cite{ADAMS2005102,ARSENE20051,PhysRevLett.105.252301,Niida2021}. The experimental signatures of QGP suggest that its initial stage is of very high energy density and low viscosity often referred to as nearly perfect fluid~\cite{Bernhard2019}. The successful explanation of experimental observables such as the elliptic flow coefficient $v_2$~\cite{PhysRevLett.87.182301, PhysRevC.77.054901, PhysRevC.78.034915} created confidence in the applicability of relativistic dissipative hydrodynamics for QGP evolution.
 The equation of state, as well as dissipative, non-equilibrium processes, influence the space-time evolution of a medium. The temperature drops when QGP expands. Hydrodynamic simulations can explain the temperature evolution very well. First-order dissipative hydrodynamic theory given by Eckart~\cite{Eckart:1940te}, and Landau and Lifshitz~\cite{landau1987fluid1} leads to an acausality problem. This problem vanishes in second-order hydrodynamics given by Muller, Israel, and Stewart, also known as MIS theory~\cite{Israel:1979wp}. 
 
The hydrodynamic stage of the QGP expansion is described as an adiabatic thermodynamic process but not strictly adiabatic~\cite{BREWER2021136189}. During the early stages of the expansion, thermal conductivity can play a vital role in the equilibration of the system. As the medium expands, the temperature drops, which can lead to local temperature gradients. This can lead to the creation of hotspots in the QGP, as shown in Ref.~\cite{Gyulassy:1996br}; hence, heat current will be generated, and the thermal conductivity of the system takes a significant role here. The thermal conductivity of quark and hadronic matter is widely investigated in microscopic theories, where two formalisms are widely used. One is Kubo formalism~\cite{PhysRevD.95.114021, Fernandez-Fraile:2009eug} based on quantum field theory, and the other is the relaxation time approximation (RTA) formalism based on kinetic theory approach~\cite{Gavin:1985ph, PhysRevD.100.114004, PhysRevD.102.016016, PhysRevD.106.034008, PhysRevD.108.094007}. In all the previous calculations, a static picture of QGP is considered, \i.e., the effect of cooling due to the expansion of QGP is not considered. In principle, cooling of QGP with time should affect the temperature gradients in the system as well, depending on the change in thermal conductivity of the medium. To account for this effect, heat current modifies depending on the rate of change of temperature gradient and change in thermal conductivity. Here, we will show that this effect on thermal conductivity is significant and non-negligible.  

According to the theoretical predictions~\cite{Tuchin2013ie}, a substantial magnetic field is created in the non-central heavy ion collision experiments, which magnetize the created medium~\cite{Wang:2021eud}. With the finite electrical conductivity of QGP, the magnetic field can survive until the medium exists~\cite {Dey:2021fbo}. The decay of the magnetic field is influenced by various parameters, such as electrical conductivity, expansion rate or cooling rate, etc. In reality, electrical conductivity changes as the medium evolves, affecting the electromagnetic field in the medium. Indeed the actual picture is very complicated, and one possible solution to a time-dependent electromagnetic field could be to solve Maxwell's equation in a self-consistent manner~\cite{Tuchin:2015oka, PhysRevD.90.034018}. Direct quantitative measurement of a magnetic field from the experiment is yet to be discovered; however, a recent discovery of separation of directed flow coefficient $v_1$ of $D^0$ and $\bar{D}^0$ proves the existence of a very high magnetic field at the early stage of QGP creation~\cite{PhysRevLett.125.022301, PhysRevLett.123.162301}. Many theoretical studies~\cite{PhysRevD.106.014011, Hongo:2013cqa, PhysRevD.106.034008} attempted to explain this experimental finding using various decay profiles of the electromagnetic field. It is found that the results are sensitive to the decay parameter and the decay profile of the magnetic field, therefore, on the electromagnetic properties of the matter. The other transport properties of QGP, such as viscosity, thermal conductivity, etc., are also highly affected by the magnetic fields~\cite{Rath2022}. Recently, a few studies~\cite{PhysRevD.106.034008, PhysRevD.104.094037} show that the time evolution of electromagnetic field has a non-negligible effect on the transport properties - thermal and electrical conductivity of QGP and hadronic matter. However, in all these estimations, a static picture of QGP is considered, \i.e., the effect of cooling is not considered, which can significantly impact transport coefficients, especially in thermal conductivity, as we will show here. 

In this work, we estimated the thermal conductivity of the evolving quark-gluon plasma in a time-varying magnetic field at finite baryon chemical potential. Here we used a quasiparticle-based model of QGP for quantitative estimations. Moreover, we investigated the deviation of the elliptic flow coefficient $v_2$ of the QGP in evolving pictures from the static one in the presence of a time-varying magnetic field. 
Furthermore, we extend the study for a (1+1)D hydrodynamical system that is longitudinally boost-invariant and has an azimuthally symmetric radial expansion with a finite viscosity as described in Gubser flow~\cite{PhysRevD.82.085027}.
The paper is organized in the following manner. The derivation of thermal conductivity components for static and evolving medium in the presence of a time-varying field is briefly given in the formalism section (\ref{sec:Formalism}). In section (\ref{sec-result}), we discuss the results in detail. In the first part of the results, we discuss various cooling rates. This is followed by thermal conductivity and the effect of different cooling rates. Then, we discuss the impact of the evolution picture on the elliptic flow coefficient. 
Further, we also provide an extended study to incorporate the effect of evolution in conductivity with the Gubser hydrodynamic model.
Finally, in section (\ref{sec-summary}), we have summarized the study with possible outlooks. Detailed conductivity and cooling rate calculations are given in the appendices.

\section{Formalism}
	\label{sec:Formalism}
In this section, we calculate the thermal conductivity of an evolving relativistic fluid in the presence of an external time-dependent magnetic field. The detailed calculations are mentioned in appendix \ref{appendix2}. 

The conservation laws associated with energy-momentum tensor $T^{\mu\nu}$ and particle-four flow $N^{\mu}$ can be used to study the fluid properties of a medium. Within the kinetic theory formalism, one can express these quantities in terms of the particle's energy, momentum, and phase space integration as
\begin{align}\label{1.1}
T^{\mu\nu} &= \sum_{i}g_i\int{\frac{d^3|\vk_i|}{(2\pi)^3}\,\frac{{k}_i^{\mu} {k}_i^{\nu}}{\om_i}\,f_i},\nn\\
N^{\mu} &= \sum_{i}g_i\int{\frac{d^3|\vk_i|}{(2\pi)^3}\,\frac{{k}_i^{\mu}}{\om_i}\,f_i}.
\end{align}
Where $k^\mu_i = (\om_i, \vec{k_i})$ is particle's four-momentum of i$^{\rm th}$ species. The total single-particle distribution function ($f_i$) for a system slightly out of equilibrium ($\delta f_{i}$) can be written as $f_i=f^0_i+\delta f_i$, where $f^0_{i}$ is total single-particle distribution function for i$^{\rm th}$ species at equilibrium, which is given by
\begin{align}\label{Dis-f}
f^0_i= \frac{1}{e^{\frac{\om_i-{\rm b}_i\mu_B}{T}}\pm 1}~,
\end{align}
where $\pm$ stands for fermion and boson, respectively. For a system slightly out of equilibrium, $T^{\mu\nu}$ and $N^{\mu}$ can be expressed as the sum of the ideal and dissipative parts, respectively as
\begin{align}\label{1.2}
    T^{\mu\nu} = T^{\mu\nu}_{\rm ideal} + \delta T^{\mu\nu}, \nn\\
    N^{\mu} = N^{\mu}_{\rm ideal} + \delta N^{\mu}.
\end{align}
The equilibrium distribution function $f_0$ contributes to the ideal parts, whereas the deviated part of the distribution function $\delta f$ contributes to the dissipative part.
Any dissipative current must have a conserved quantity, which in the case of heat current in QGP is baryon number b$_i$. As a result, only quarks (and antiquarks) contribute to thermal conduction in QGP. When the total number of gluons is conserved, gluons can contribute to the heat conductivity of a system. According to Ref.~\cite{Chakrabarty:1985ikw}, the contribution is negligible (about $10^7$ times smaller) compared to the quark-antiquark mixture. Therefore, we ignore the thermal conductivity caused by gluons in our work, yet they contribute to the system's total enthalpy and affect its thermal conductivity.
In general heat current in terms of $T^{\mu\nu}$ and $N^{\mu}$ can be expressed as~\cite{Gavin:1985ph}
\begin{align}\label{1.4}
 I^j= \delta T^{0j} -h\, \delta N^j,  \end{align}
where $h=\frac{\ep + P}{n}$ is the enthalpy per particle, $\ep$, $P$, and $n$ are total energy density, total pressure, and net baryon density of the system, respectively. 
Employing Eq.~(\ref{1.1}) in Eq.~(\ref{1.4}), we can express the heat current due to conserved baryon number in the kinetic theory for i$^{\rm th}$ species particles as, 
\begin{align}\label{1.5}
{\vec {I}}_i = \int \frac{d^3|\vk_i|}{(2\pi)^3} \frac{\vec {k}_i}{\om_i} (\om_i -{\rm b}_i h)\delta f_i.
\end{align}
To find the expression of $\delta f_i$, we solve the Boltzmann transport equation (BTE) using relaxation time approximation (RTA). 
RTA, to satisfy energy-momentum conservation law, we have to work with Landau-Lifshitz frame~\cite{landau1987fluid1}, where the rest frame is attached to energy flow. In the local rest frame, heat current will take the form,~\cite{Hosoya:1983xm, PhysRevD.106.034008}, 
\begin{align}\label{Thermcond}
	\vec{I} &= -\kappa_0  \vec{\nabla}T~,   
\end{align}
where $\kappa_0$ is the coefficient of thermal conductivity.
In the current work, we introduce the cooling rate using hydro-dynamical theories and study the thermal response of evolving QGP. 
Now we discuss the static and the evolving pictures of QGP one by one. 

\subsection{Static picture (without temperature evolution)}
\label{subsec2A}
In this picture, we discuss two cases in the presence and absence of an external (time-varying) magnetic field to study the thermal response of the medium. 
In the presence of a time-varying magnetic field, heat current in the fluid rest frame can be expressed as~\cite{PhysRevD.106.034008}, 
\begin{align}\label{I1}
	\vec{I} &= -\big\{\kappa_0^s  \vec{\nabla}T +\bar{\kappa}_1^s (\vec{\nabla}T \times \vec{B}) +\bar{\kappa}_2^s (\vec{\nabla}T \times \dot{\vec{B}})\big\}\nonumber \\ 
	 &=-\big\{\kappa_0^s \vec{\nabla}T +  (\kappa_1^s+\kappa_2^s) (\vec{\nabla}T \times \hat{b})\big\}\nonumber \\ 
	 &= -\big\{\kappa_0^{s} \vec{\nabla}T +  \kappa_H^{s} (\vec{\nabla}T \times \hat{b})\big\}~.   
\end{align}
$\kappa_0^s$ is the components of thermal conductivity along the temperature gradient, and the Hall-like components are $\kappa_1^s=\bar{\kappa}_1^s {B}$, $\kappa_2^s =\bar{\kappa}_2^s \dot{{B}}$. Here, superscript $s$ in $\kappa^s$ corresponds to a static picture.

Here, Eq.~(\ref{1.5}) also expresses the microscopic three-vector form of heat current. To find the expression of $\delta f_i$, we solve the BTE in the presence of an external magnetic field under the RTA as
\begin{align}\label{BTE-RTA-Th}
	\frac{\del f_i}{\del \tau} + \frac{\vk_i}{\om_i} \cdot \frac{\del f_i}{\del \vec{x}} + q_i \left(\frac{\vk_i}{\om_i} \times \vec{B}\right) \cdot \frac{\del f_i}{\del \vec{k_i}}
	= -\frac{\delta f_i}{\tau^i_R}~.
\end{align}
We consider a time-varying magnetic field of the form~\cite{PhysRevD.90.034018, Hongo:2013cqa}
\begin{align}\label{Magfield}
	B = B_0 \exp{\left(-\frac{\tau}{\tau_B}\right)},
\end{align}
where $B_0$ is magnitude of the initial field having decay parameter  $\tau_B$, and $\tau$ is the proper time.
We can assume an ansatz of $\delta f_i$ as \cite{Gavin:1985ph}
\begin{equation}\label{1.6}
\delta f_i=\frac{({\vec{k_i}}.{\vec \Omega}_\kappa )}{T} \frac{\partial f^0_i}{\partial \om_i}, 
\end{equation} 
where a general form of ${\vec{\Omega}_\kappa}$ up to first order time derivative of $\vec B$ can be expressed as
\begin{align}\label{1.7}
\vec{\Omega}_\kappa = \alpha_1\vec{B}+ \alpha_2\vec{\nabla}T+ \alpha_3(\vec{\nabla}T \times \vec{B})+\alpha_4 \dot{\vec{B}} +\alpha_5(\vec{\nabla}T \times \dot{\vec{B}}).
\end{align} 
Considering the same magnetic field profile as mentioned in  Eq.~(\ref{Magfield}), we can find the unknown coefficients $\alpha_{i}$ ($i=(1, 2,.., 5)$) by solving Eq.~(\ref{BTE-RTA-Th}) employing Eq.~(\ref{1.7}) in Eq.~(\ref{1.6}). From Eqs.~(\ref{I1}) and (\ref{1.5}), we can obtain the expressions of thermal conductivity components as \cite{PhysRevD.108.094007}

    \begin{align}\label{18}
    \kappa_0^{s} =  \frac{1}{3T^2} \sum_i g_i \int \frac{d^3|\vk_i|}{(2\pi)^3}\frac{\vec{k}^2_i}{\om_i^2}(\om_i - {\rm b}_i h)^2  \tau_R^i &\frac{1}{(1+\chi_i + \chi_i^2)}\nn\\
    &~~~f^0_i(1\mp f^0_i) ~,\nn\\
    {\kappa}_H^{s} = \frac{1}{3T^2} \sum_i g_i \int \frac{d^3|\vk_i|}{(2\pi)^3}\frac{\vec {k}^2_i}{\om_i^2}(\om_i - {\rm b}_i h)^2 \tau_R^i &\frac{\chi_i}{(1+\chi_i + \chi_i^2)}\nn\\
    &~~~f^0_i(1\mp f^0_i) ~,
    \end{align}
with, $\chi_i = \frac{\tau_R^i}{\tau_B}$. 

The methodology adopted here is the same as Refs.~\cite{PhysRevC.93.014903, PhysRevD.94.094002} in the absence of an external magnetic field, \i.e. ($B = 0$). Unlike the case of the presence of a magnetic field, there is only a single component of the thermal conductivity as 
\begin{equation}
\kappa^{s} = \frac{1}{3T^2} \sum_i g_i \int \frac{d^3|\vk_i|}{(2\pi)^3}\frac{\vec{k}^2_i}{\om_i^2}(\om_i - {\rm b}_i h)^2 \tau_R^i ~f^0_i(1\mp f^0_i)~.
\label{th-B0}
\end{equation}

\subsection{Evolving picture (with temperature evolution)}
In the evolution picture, we consider the evolution of QGP using temperature evolution with time, for which we have taken different hydrodynamics cooling rates, which are discussed below in three cases. 

In an evolving system, the Eq.~(\ref{Thermcond}) gets modified. As the temperature falls with time, the temperature gradient ($\vec\nabla T$) should also change, which will have a finite effect on heat current.
Therefore, in addition to $\vec\nabla T$, there are also contributions $\vec{\nabla} \dot{T}$ in the heat current $\vec I$. Hence, in the evolution picture, the heat current in the rest frame of the fluid can be expressed as
\begin{align}\label{Theat}
	\vec{I} &= -(\kappa_0 \vec{\nabla}T + \Bar{\kappa}_1 \vec{\nabla}\dot T)  ~,   
\end{align}
where $\Bar{\kappa}_1$ is a new coefficient of thermal conductivity arising after introducing cooling rate or $\vec\nabla \dot T$ term.
Similarly, in the presence of (time-varying) magnetic field heat current Eq.~(\ref{I1}) will also be modified for evolving picture as~
\begin{align}\label{IIc}
\vec{I} &= -\big\{\kappa_0 \vec{\nabla}T +\bar{\kappa}_1 (\vec{\nabla}T \times \vec{B}) +\bar{\kappa}_2 (\vec{\nabla}T \times \dot{\vec{B}}) + \bar{\kappa}_3  \vec{\nabla}\dot T \nonumber\\
 &+\bar{\kappa}_4 (\vec{\nabla}\dot T \times\vec{B}) \big\}~,   
\end{align} 
where, $\bar{\kappa}_3$ and $\bar{\kappa}_4$ are the new coefficients arising due to temperature evolution. All these new components are calculated below for three cooling rates obtained from ideal hydrodynamics, ideal magnetohydrodynamics, and first-order dissipative hydrodynamics.

\subsubsection{Case-I: Ideal hydrodynamics ($B = 0$)}
The expression of cooling rate in ideal hydrodynamics with the parameters, medium formation time $\tau = \tau_{0}$ and initial temperature $T = T_{0}$ is (see Appendix~\ref{appendix1} for more detail)
\begin{align} \label{ITemp}
T = T_{0} {\left(\frac{\tau_{0}}{\tau}\right)}^{\frac{1}{3}}.
\end{align}

There is no dissipation or dissipative current in ideal hydrodynamics, whereas conductivity is a dissipative quantity. However, here we use ideal hydro only for the cooling rate, and the calculation of conductivity using this cooling rate will only give us a reference. Therefore, the cooling rate from dissipative hydro is a must for actual conductivity measurement, and we have used only first-order hydro in case-III. In the case of evolving medium in the fluid rest frame, the heat current can be expressed as mentioned in Eq.~(\ref{Theat})
\begin{align}\label{III}
\vec{I} & = -\kappa^{e} \vec{\nabla}T~.   
\end{align} 
Where, $\kappa^e$ = $\kappa_0 + \kappa_1$ and $\kappa_1$ = $\frac{-1}{3\tau}\Bar{\kappa}_1$. The superscript $e$ in $\kappa^e$ corresponds to the evolving picture.
To obtain a microscopic expression of heat current using the cooling rate Eq.~(\ref{ITemp}), we follow a similar prescription as discussed in \ref{subsec2A}. We find the deviated part of the distribution function $\delta f$ by solving the BTE for evolving medium under the RTA as
\begin{align}\label{1.9A}
	\frac{\del f_i}{\del \tau} + \frac{\vk_i}{\om_i} \cdot \frac{\del f_i}{\del \vec{x}}
	= -\frac{\delta f_i}{\tau^i_R}~.
\end{align}
We can assume an ansatz of $\delta f_i$ for thermal conductivity as \cite{Gavin:1985ph}
\begin{equation}\label{ans1}
\delta f_i=\frac{({\vec{k_i}}.{\vec \Omega}_\kappa )}{T} \frac{\partial f^0_i}{\partial \om_i}, 
\end{equation} 
where a general form of ${\vec{\Omega}_\kappa}$ up to first order time derivative of $\vec T$ can be expressed as
\begin{align}\label{1.91}
\vec{\Omega}_\kappa =&  \alpha_1\vec{\nabla}T + \alpha_2\vec{\nabla}\dot T~. 
\end{align} 
The unknown coefficients $\alpha_1$ and $\alpha_1$ can be found by substituting  Eq.~(\ref{1.91}) and Eq.~(\ref{ans1}) in Eq.~(\ref{1.9A}). Finally,  the expression of thermal conductivity for evolving medium (ideal hydrodynamic evolution) can be obtained as (see Appendix \ref{appendix2}),
\begin{widetext}
 \begin{align}\label{21}
    \kappa_0 &=  \frac{1}{3T^2} \sum_i g_i \int \frac{d^3|\vk_i|}{(2\pi)^3}\frac{\vec{k}^2_i}{\om_i^2}(\om_i - {\rm b}_i h)^2 {\tau_R^i}  f^0_i(1\mp f^0_i) 
    - \frac{1}{3T^2} \sum_i g_i \int \frac{d^3|\vk_i|}{(2\pi)^3}\frac{\vec{k}^2_i}{\om_i}(\om_i - {\rm b}_i h) {\tau_R^i}  \exp(-\tau/\tau_R) f^0_i(1\mp f^0_i),\nn\\
     {\kappa}_1 &= \frac{1}{9\tau T^2} \sum_i g_i \int \frac{d^3|\vk_i|}{(2\pi)^3}\frac{\vec{k}^2_i}{\om_i^2}(\om_i - {\rm b}_i h)^2 {\tau_R^i}^2  f^0_i(1\mp f^0_i) 
    - \frac{1}{9\tau T^2} \sum_i g_i \int \frac{d^3|\vk_i|}{(2\pi)^3}\frac{\vec{k}^2_i}{\om_i}(\om_i - {\rm b}_i h) {\tau_R^i}^2  \exp(-\tau/\tau_R) f^0_i(1\mp f^0_i),\nn\\
{\rm or,}\nn\\
    {\kappa}^e &= \kappa_0 + \kappa_1, \nn\\ 
    &= \frac{1}{3T^2} \sum_i g_i \int \frac{d^3|\vk_i|}{(2\pi)^3}\frac{\vec{k}^2_i}{\om_i^2}(\om_i - {\rm b}_i h) {\tau_R^i}\left(1 + \frac{\tau_R^i}{3\tau}\right)\left\{ (\om_i - {\rm b}_i h)- \om_i \exp(-\tau/\tau_R)\right\}  
    f^0_i(1\mp f^0_i)~. 
    \end{align}   
\end{widetext}
%

\subsubsection{Case-II: Ideal magnetohydrodynamics}
In the presence of a magnetic field, the cooling of the medium is affected. In the case of ideal magnetohydrodynamics, the cooling rate for a time-dependent magnetic field can be obtained as (see Appendix \ref{appendix1}),
\begin{align} \label{bTemp}
T &= \Big[T_{0}^4 {\left(\frac{\tau_{0}}{\tau}\right)}^{\frac{4}{3}} + \frac{4\alpha}{{(2\beta\tau)}^{\frac{4}{3}}} \Big\{\Gamma{(4/3, 2\beta\tau)} - \Gamma(4/3, 2\beta\tau_0) \Big\} \nn\\
&~~~~- \frac{2\alpha}{{(2\beta\tau)}^{\frac{7}{3}}}\Big\{\Gamma(7/3, 2\beta\tau) - \Gamma(7/3, 2\beta\tau_0) \Big\}\Big]^\frac{1}{4}.
\end{align}
Where $\tau_{0}$ and $T_{0}$ are medium formation time and initial temperature, respectively; $\alpha = \frac{B_0^2}{12a}$ with $a = \left(16 + \frac{21}{2}N_{f}\right)\frac{\pi^{2}}{90}$
and $\beta = \frac{1}{\tau_B}$. Here $N_{f}$ is the number of quark flavors, taken to be three.
In the considered evolving picture, heat current using Eq.~(\ref{bTemp}) and Eq.~(\ref{IIc}) can be expressed as
\begin{align}\label{II}
\vec{I} &= -\big\{\kappa_0 \vec{\nabla}T +\bar{\kappa}_1 (\vec{\nabla}T \times \vec{B}) +\bar{\kappa}_2 (\vec{\nabla}T \times \dot{\vec{B}}) + \bar{\kappa}_3  \vec{\nabla}\dot T \nonumber\\
 &~~~+\bar{\kappa}_4 (\vec{\nabla}\dot T \times\vec{B}) \big\}~,\nn\\ 
 &= -\big\{(\kappa_0 + \kappa_3) \vec{\nabla}T +  (\kappa_1+\kappa_2+\kappa_4) (\vec{\nabla}T \times \hat{b})\big\}\nonumber \\ 
&= -\big\{\kappa_0^{e} \vec{\nabla}T +  \kappa_H^{e} (\vec{\nabla}T \times \hat{b})\big\}~.  \end{align} 
Here, $\kappa_1=\bar{\kappa}_1 {B}$, $\kappa_2 =\bar{\kappa}_2 \dot{{B}}$, $\kappa_3=\bar{\kappa}_3 \big\{\frac{-1}{3\tau}\left(1 - \frac{3B^2}{4aT^4} \right)-\frac{B^2}{4a\tau_{B}T^4} \big\}$, and $\kappa_4=\bar{\kappa}_4\big\{\frac{-1}{3\tau}\left(1 - \frac{3B^2}{4aT^4} \right)-\frac{B^2}{4a\tau_{B}T^4} \big\} {B}$.
From Eq.~\ref{bTemp},
\begin{align}
    \vec{\nabla}\dot T =  \left\{\frac{-1}{3\tau}\left(1 - \frac{3B^2}{4aT^4} \right)-\frac{B^2}{4a\tau_{B}T^4} \right\}\vec{\nabla}T~,
\end{align} 
(see Appendix \ref{appendix1}).

To obtain the expression of conductivity components, we follow the same methodology as for the case of static picture~(\ref{subsec2A}). In this case, the ansatz $\delta f_i$ will be modified such that Eq.~(\ref{1.7}) takes the form, 
\begin{align} \label{1.7a1}
\vec{\Omega}_\kappa &= \alpha_1\vec{B}+ \alpha_2\vec{\nabla}T+ \alpha_3(\vec{\nabla}T \times \vec{B})+\alpha_4 \dot{\vec{B}} +\alpha_5(\vec{\nabla}T \times \dot{\vec{B}})\nn\\
 &+\alpha_6\vec{\nabla}\dot T+\alpha_7(\vec{\nabla}\dot T \times \vec{B})+\alpha_8(\vec{\nabla}\dot T \times \dot{\vec{B}}),
\end{align} 
where we considered up to first order time derivative of $\vec B$ and $\vec{\nabla} T$.
The unknown coefficients $\alpha_{i}$ ($i=(1, 2.....8)$) can be obtained by substituting  Eq.~(\ref{1.7a1}) and Eq.~(\ref{1.6}) in Eq.~(\ref{BTE-RTA-Th}). Solving the Boltzmann equation, we can obtain the expressions of thermal conductivity components from Eqs.~(\ref{1.5}) and (\ref{II}) as (see Appendix \ref{appendix2}).
\begin{widetext}
    \begin{align}\label{25}
    \kappa_0 &=  \frac{1}{3T^2} \sum_i g_i \int \frac{d^3|\vk_i|}{(2\pi)^3}\frac{\vec{k}^2_i}{\om_i^2}(\om_i - {\rm b}_i h)^2 \tau_R^i ~\frac{1}{(1+\chi_i + \chi_i^2)}~ f^0_i(1\mp f^0_i),\nn\\
     {\kappa}_1 &= \frac{1}{3T^2} \sum_i g_i \int \frac{d^3|\vk_i|}{(2\pi)^3}\frac{ \vec{k}^2_i}{\om_i^2}(\om_i - {\rm b}_i h)^2 \tau_R^i ~\frac{\chi_i}{(1+\chi_i)(1+\chi_i + \chi_i^2)} ~f^0_i(1\mp f^0_i), \nn\\
    {\kappa}_2 &= \frac{1}{3T^2} \sum_i g_i \int \frac{d^3|\vk_i|}{(2\pi)^3}\frac{\vec {k}^2_i}{\om_i^2}(\om_i - {\rm b}_i h)^2 \tau_R^i ~\frac{\chi_i^2}{(1+\chi_i)(1+\chi_i + \chi_i^2)} ~f^0_i(1\mp f^0_i), \nn\\
\kappa_3 &=  \frac{1}{3T^2} \sum_i g_i \int \frac{d^3|\vk_i|}{(2\pi)^3}\frac{\vec{k}^2_i}{\om_i^2}(\om_i - {\rm b}_i h)^2 \tau_R^{i2} \frac{\left((1+\chi_i)(\chi_i^2-1)\right) + \chi_i^2}{(1+\chi_i)(1+\chi_i + \chi_i^2)(1+\chi_i^2)}
\left\{\frac{-1}{3\tau}\left(1 - \frac{3B^2}{4aT^4} \right)-\frac{B^2}{4a\tau_{B}T^4} \right\}~ f^0_i(1\mp f^0_i),\nn\\
\kappa_4 &=  \frac{1}{3T^2} \sum_i g_i \int \frac{d^3|\vk_i|}{(2\pi)^3}\frac{\vec{k}^2_i}{\om_i^2}(\om_i - {\rm b}_i h)^2\tau_R^{i2} \frac{\chi_i + \chi_i(1+\chi_i)}{(1+\chi_i)(1+\chi_i + \chi_i^2) (1+ \chi_i^2)}\left\{\frac{1}{3\tau}\left(1 - \frac{3B^2}{4aT^4} \right)+\frac{B^2}{4a\tau_{B}T^4} \right\}~ f^0_i(1\mp f^0_i),\nn\\
{\rm or,}\nn\\
  {\kappa}_0^{e} &= \kappa_0 + \kappa_3, \nn\\
  &= \frac{1}{3T^2} \sum_i g_i \int \frac{d^3|\vk_i|}{(2\pi)^3}\frac{\vec{k}^2_i}{\om_i^2}(\om_i - {\rm b}_i h)^2 \frac{\tau_R^i} {(1+\chi_i + \chi_i^2)}
  \left[1 - \left\{{\tau_R^{i}} \frac{(1+\chi_i)(\chi_i^2-1) + \chi_i^2}{(1+\chi_i)(1+\chi_i^2)}\right\}
\left\{\frac{1}{3\tau}\left(1 - \frac{3B^2}{4aT^4} \right)+\frac{B^2}{4a\tau_{B}T^4}\right\}\right]\nn\\
&\hspace{149mm} \times f^0_i(1\mp f^0_i),  \nn\\
{\kappa}_H^{e} &= \kappa_1 + \kappa_2 + \kappa_4, \nn\\
&= \frac{1}{3T^2} \sum_i g_i \int \frac{d^3|\vk_i|}{(2\pi)^3}\frac{\vec{k}^2_i}{\om_i^2}(\om_i - {\rm b}_i h)^2 \frac{\tau_R^i \chi_i} {(1+\chi_i + \chi_i^2)}
  \left[1 + \left\{{\tau_R^{i}} \frac{(2+\chi_i)}{(1 + \chi_i)(1+\chi_i^2)}\right\}
\left\{\frac{1}{3\tau}\left(1 - \frac{3B^2}{4aT^4} \right)+\frac{B^2}{4a\tau_{B}T^4}\right\}\right]\nn\\
&\hspace{140mm} \times f^0_i(1\mp f^0_i).  
\end{align}
\end{widetext}
with $\chi_i = \frac{\tau_R^i}{\tau_B}$.
Here, it is essential to note that the expressions are obtained in the limit of slowly varying magnetic field, for which we approximated the decay parameter as the inverse of cyclotron frequency, \i.e., $\tau_B = \frac{\om_i}{q_iB}$. Furthermore, $\kappa_H$ should have explicit sign dependency from the electric charge of the particle due to $\chi_i$ on the numerator. However, this information vanishes due to the approximation, leading to inaccurate results. Therefore, we use the minus (plus) sign in $\kappa_H$ for negatively (positively) charged particles and antiparticles for the numerical estimations.

\subsubsection{Case-III: First-order dissipative hydrodynamics ($B = 0$)}
The expression of cooling rate in the first-order dissipative theory with the parameters, medium formation time $\tau = \tau_{0}$ and initial temperature $T = T_{0}$ can be obtained as mentioned in ref.~\cite{Muronga:2001zk} 
\begin{align} \label{FTemp}
T = T_{0} {\left(\frac{\tau_{0}}{\tau}\right)}^{\frac{1}{3}}\left\{1 + \frac{b}{6a}\frac{1}{\tau_0 T_0}\left(1 - (\frac{\tau_{0}}{\tau})^{\frac{2}{3}}\right)\right\}~,
\end{align}
where $a$ is same as used in Eq.~(\ref{bTemp}), and 
\begin{equation}
b=\left(1 + 1.70N_{f}\right)\frac{0.342}{(1+N_{f}/6)\alpha_s^{2}\ln(\alpha_s^{-1})}.\nn\\   
\end{equation}
Here, $\alpha_s$ is the strong fine structure constant and taken to be 0.5, and $N_f$ is the number of quark flavors, taken to be three.
To get the final expression for thermal conductivity, the formalism followed here for this case is similar to what we followed for case-I. The difference in both cases comes from their respective cooling rates. Instead of Eq.~(\ref{ITemp}), here we use the cooling rate from Eq.~(\ref{FTemp}) into Eq.~\ref{III}. 
Hence, the obtained expression of thermal conductivity components for this case is
\begin{widetext}
 \begin{align}\label{27}
    \kappa_0 &=  \frac{1}{3T^2} \sum_i g_i \int \frac{d^3|\vk_i|}{(2\pi)^3}\frac{\vec{k}^2_i}{\om_i^2}(\om_i - {\rm b}_i h)^2 {\tau_R^i}  f^0_i(1\mp f^0_i) 
    - \frac{1}{3T^2} \sum_i g_i \int \frac{d^3|\vk_i|}{(2\pi)^3}\frac{\vec{k}^2_i}{\om_i}(\om_i - {\rm b}_i h) {\tau_R^i}  \exp(-\tau/\tau_R) f^0_i(1\mp f^0_i),\nn\\
     {\kappa}_1 &= \frac{1}{3\tau T^2} \sum_i g_i \int \frac{d^3|\vk_i|}{(2\pi)^3}\frac{\vec{k}^2_i}{\om_i^2}(\om_i - {\rm b}_i h)^2 {\tau_R^i}^2  f^0_i(1\mp f^0_i) 
    - \frac{1}{3\tau T^2} \sum_i g_i \int \frac{d^3|\vk_i|}{(2\pi)^3}\frac{\vec{k}^2_i}{\om_i}(\om_i - {\rm b}_i h) {\tau_R^i}^2  \exp(-\tau/\tau_R) f^0_i(1\mp f^0_i),\nn\\
{\rm or,}\nn\\
    {\kappa}^e &= \kappa_0 + \kappa_1, \nn\\ 
    &= \frac{1}{3T^2} \sum_i g_i \int \frac{d^3|\vk_i|}{(2\pi)^3}\frac{\vec{k}^2_i}{\om_i^2}(\om_i - {\rm b}_i h) {\tau_R^i}\left(1 + \frac{\tau_R^i}{\tau}\right)\left\{ (\om_i - {\rm b}_i h)- \om_i \exp(-\tau/\tau_R)\right\} 
    f^0_i(1\mp f^0_i). 
\end{align}   
\end{widetext}
Where, $\kappa_1$ = $\frac{-1}{\tau}\Bar{\kappa}_1$.
To satisfy the energy-momentum conservation law in the RTA method, the relaxation time must be momentum-independent, which is not the case in general. However, we can find a momentum-independent relaxation time of quark in QCD matter as given in Ref.~\cite{Hosoya:1983xm}
\begin{align}
    \tau_R = \frac{1}{5.1T\alpha_s^{2}\log(1/\alpha_s)[1+0.12(2N_f+1)]}.
\end{align}
Here, we considered a fixed value of the strong coupling constant $\alpha_s = 0.5$. 

\section{Result and discussions}
\label{sec-result}
\begin{figure}
	\centering
	\includegraphics[scale=0.42]{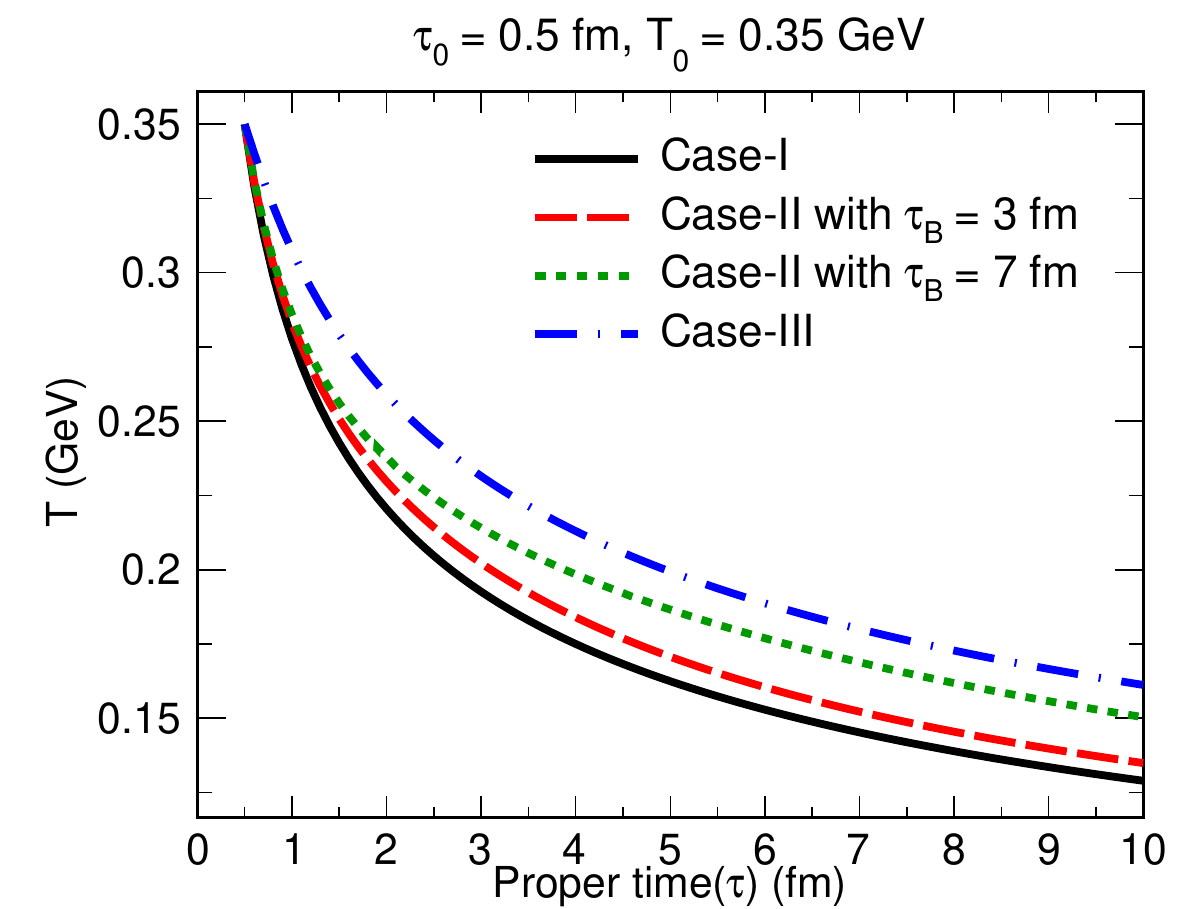}
	\caption{Cooling rate for three cases. Case-I: Ideal hydrodynamics, Case-II: Ideal magnetohydrodynamics, Case-III: First-order hydrodynamics.}
	\label{Fig-temp}
\end{figure}
In Fig.~\ref{Fig-temp}, we have plotted the cooling rate for the QGP medium (temperature ($T$) versus proper time ($\tau$)) for three cases. Case-I is for ideal hydrodynamics (solid black line), case-II is for ideal magnetohydrodynamics (red dashed for $\tau_B$ = 3 fm and green dotted line for $\tau_B$ = 7 fm), and case-III is for first-order dissipative hydrodynamics without magnetic field (blue dash-dot line). For all the cases, we have considered the same formation time ($\tau_0$) and thermalization temperature ($T_0$) of QGP as $\tau_0 = 0.5$~fm and $T_0 = 0.35$~GeV, respectively. Expression of cooling rate for the corresponding cases is given by Eqs.~(\ref{ITemp}), (\ref{bTemp}), and (\ref{FTemp}), respectively.
In case-II, we have considered a time-dependent magnetic field with decay parameter $\tau_B = 3$~fm indicated by the red dashed line and $\tau_B = 7$~fm by the green dotted line. The smaller the decay parameter, the faster the decay of the magnetic field in the medium. The strength of the initial magnetic field depends on the collision energy, impact parameter, system size, etc. Here, we have considered the initial magnetic field $eB_0 = 9 ~m_\pi^2$, which is possible at RHIC and LHC energies~\cite{Tuchin2013ie}.
From the figure, it is also noticeable that at the initial time of temperature evolution, the case-I and case-II cooling rates are the same, almost up to $2$~fm. 
Including the dissipation or the magnetic field, the cooling rate slows down compared to ideal hydrodynamics. Furthermore, slower decay of the magnetic field leads to slower cooling of the medium. In the coming discussion, we will see the effect of these different cooling rates on the thermal conductivity of the medium and, hence, its related phenomenological quantity. 

In this work, for the numerical estimations, we used a quasiparticle model formulated by Gorenstein and Yang~\cite{Gorenstein:1995vm}, where the lattice equation of state for QGP is achieved by introducing the thermal mass of the partons. By virtue of interactions, quarks and gluons get their thermal mass $m(T)$, and the thermodynamic consistency is achieved by introducing bag constant arising from vacuum energy~\cite{Srivastava:2010xa}.  
The dispersion relation of the particle in the quasiparticle model having energy $\om_i$ and momentum $k_i$ is $\om_i^{2}(k_i, T) = k_i^{2} + m_i^{2}(T)$ ~\cite{Gorenstein:1995vm}. For further details of quasiparticle models, one can check the references~\cite{Gorenstein:1995vm, Srivastava:2010xa, Bannur:2006hp, Peshier:1995ty, Peshier:1999ww}.

%
\subsection{Thermal conductivity}
\begin{figure*}
	\centering
	\includegraphics[scale=0.43]{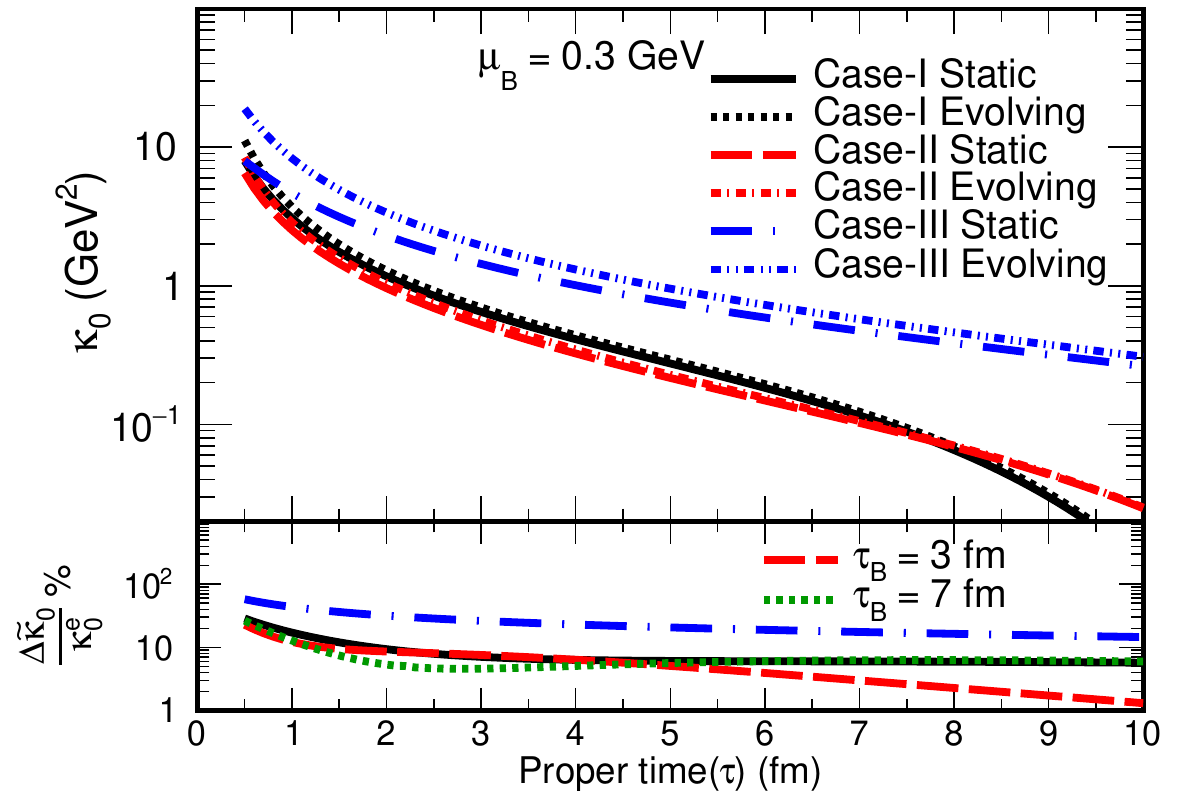}
     \includegraphics[scale=0.43]{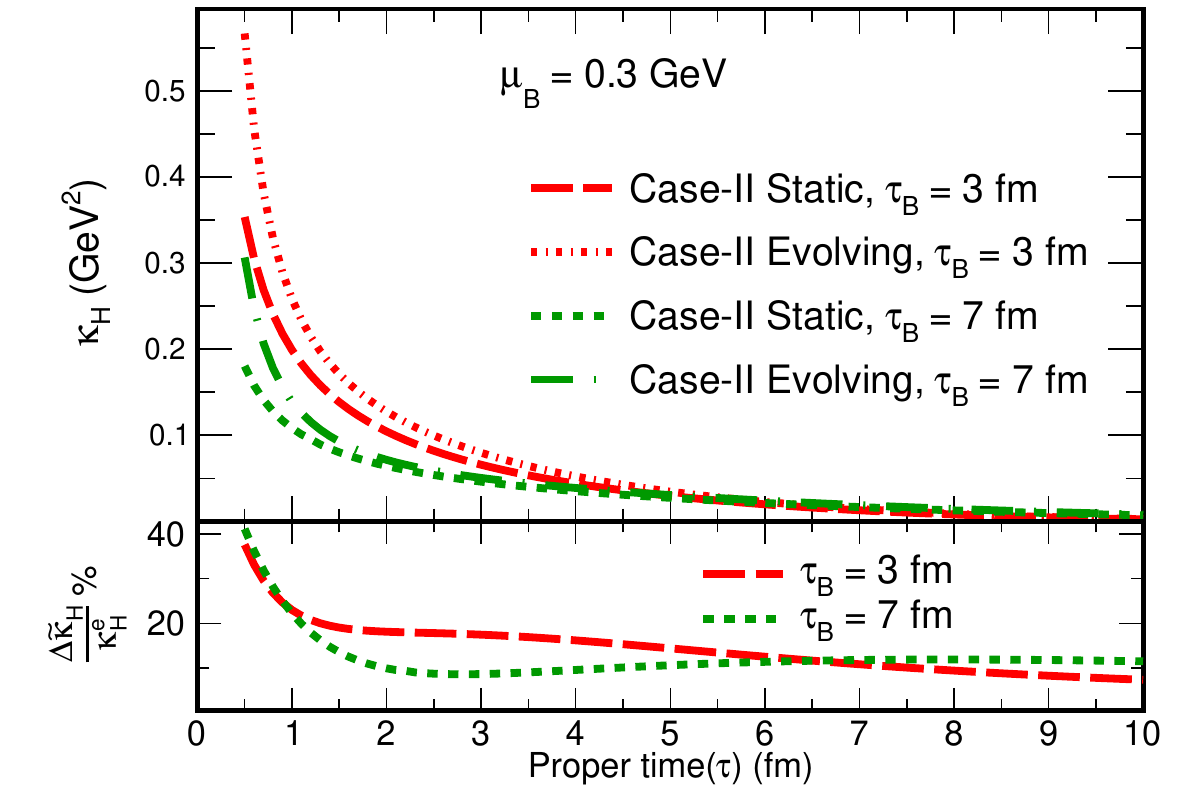}
	\caption{Components of thermal conductivity as a function of proper time ($\tau$). Left: upper panel represents $\kappa_0$ component, bottom panel represents its percentage deviation (where $\Delta\Tilde\kappa_0 = \kappa_0^{e} - \kappa_0^{s}$) of static picture from evolving picture. Right: same as the left figure but for``Hall-like" component.}
	\label{Fig-thermohmtime}
\end{figure*}
In Fig.~\ref{Fig-thermohmtime} (left), the top panel represents $\kappa_0$ component of thermal conductivity with proper time ($\tau$) at baryon chemical potential $\mu_{B}$ = 0.3 GeV for all the three cases. This component is responsible for the heat current along the temperature gradient. When the effect of cooling is considered in the conductivity calculation, the corresponding result for each case is denoted by the term ``evolving," and where it is not considered is designated by ``static." Black lines represent case-I with a dotted line for the evolving picture and a solid line for the static one. Case-II represents a red dashed line for a static picture and a red dash-single dot line for the evolving one. The blue lines represent case-III with a dash-double dotted line for the evolving picture and a dash-dot line for the static one.
In all three cases, we see an enhancement in the thermal conductivity in the evolving picture as compared to the static one. This comes from the new components $\kappa_1$ in case-I and case-III, and $\kappa_3$ in case-II arising due to $\vec{\nabla} \dot{T}$ term in heat current. 

In static pictures, the expression of thermal conductivity (Eq.~(\ref{th-B0})) is the same for case-I and case-III except for their temperature evolution. Hence, we can see that the solid black line and blue dash-dotted line start from the same point but later on follow different paths due to their respective cooling rates. For all three cases, we notice that conductivity decreases with time, mainly dominated by their thermodynamical phase space part for baryons as expected from the existing knowledge~\cite{Gavin:1985ph}. The effect of the magnetic field can be found in further suppression of conductivity, as introduced in case-II. We can quantify the effect of the cooling rate and magnetic field by looking at their respective expressions of conductivity. In the absence of magnetic field, comparing the static picture (Eq.~\ref{th-B0}) with evolving pictures (Eqs.(\ref{21}) and (\ref{27})), we can argue that the thermodynamical phase space part $(\om_i - {\rm b}_ih)$ as well as the relaxation time $\tau_R^i$ get modified in the evolving picture. However, in the presence of a magnetic field comparing static picture (Eq.(\ref{18})) and evolving picture (\ref{25}), one observes the effect of evolution is mainly taken care of by relaxation time or effective relaxation time. 
From the bottom panel of Fig.~(\ref{Fig-thermohmtime}) (left), we can quantify the effect of cooling for all the cases. On the average, $10\% - 20\%$ effects are there for all the cases. In the presence of a magnetic field, the effect is comparatively less and decreases with increasing decay parameters.    
The effect of evolution is greater in case-III than in other cases, which indicates that the evolution of thermal conductivity is highly sensitive to the cooling rate of the system.
We can roughly state that the slower the cooling higher the effect of evolution in the conductivity, and the effect decreases with time evolution.

In Fig.~\ref{Fig-thermohmtime} (right), the time evolution of the ``Hall-like" component and the effect of evolution in this component is demonstrated at baryon chemical potential $\mu_{B} = 0.3$~GeV. This component arises only due to the magnetic field, and the heat current for this component is perpendicular to both the magnetic field and temperature gradient. This figure represents $\kappa_H$ for case-II at two decay parameters, $\tau_B = 3,~7$~fm. The absolute value of $\kappa_H$ is almost ten times less than that of the $\kappa_0$ component because the contribution from particles and antiparticles are opposite, leading to a smaller net value. 
Similar to the $\kappa_0$ component, we see an enhancement in $\kappa_H$ in the evolving picture as compared to the static one. In this case, enhancement comes from the new component $\kappa_4$ arising in the evolution picture due to $(\vec{\nabla} \dot T \times \vec B) $ term.
In the bottom panel of the figure, we notice that the effect of evolution is large in the early stages of evolution (around $40\%$) as compared to the later stages (around $10\%$). As the system evolves with time, the strength of the magnetic field also starts to weaken, and the $\kappa_H$ component diminishes, leaving only the $\kappa_0$ component.
\begin{figure*}
	\centering
	\includegraphics[scale=0.40]{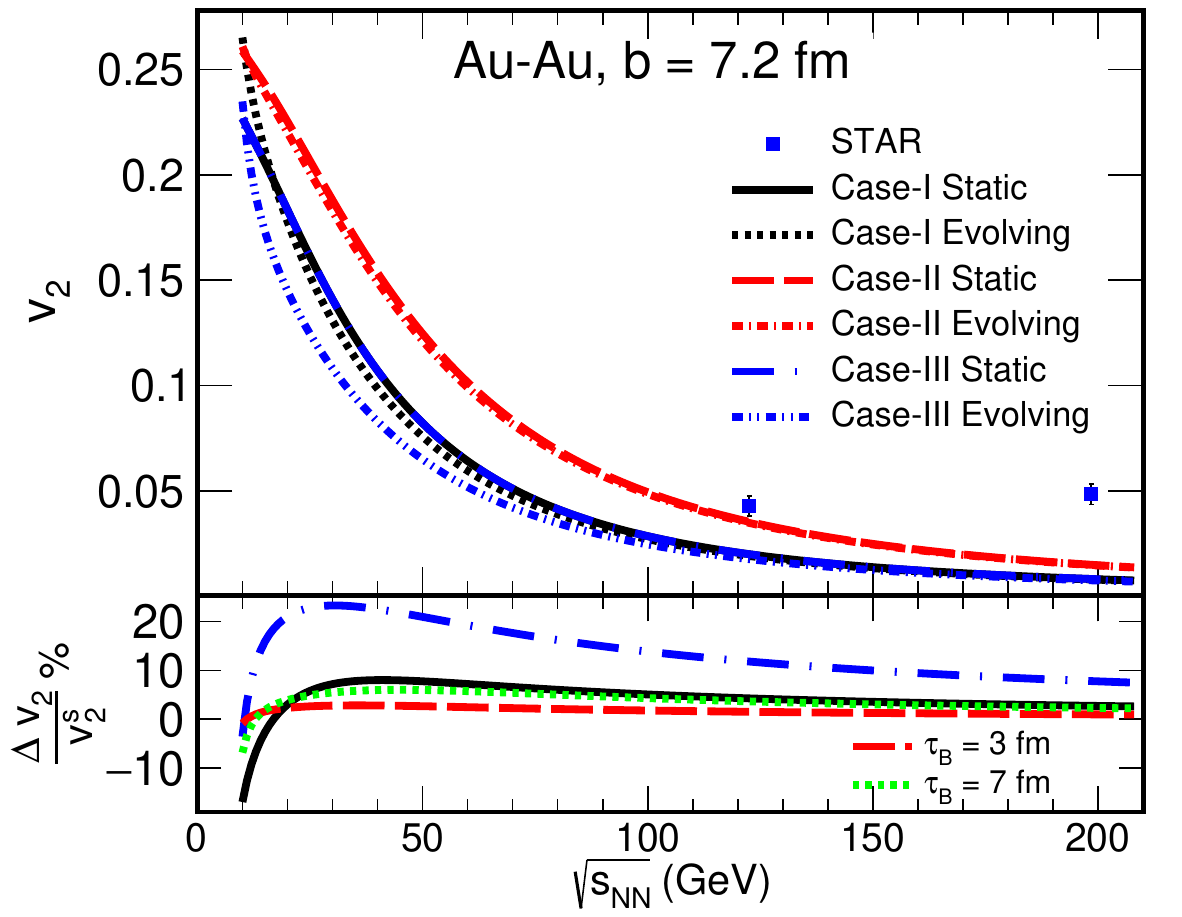}
 \includegraphics[scale=0.44]{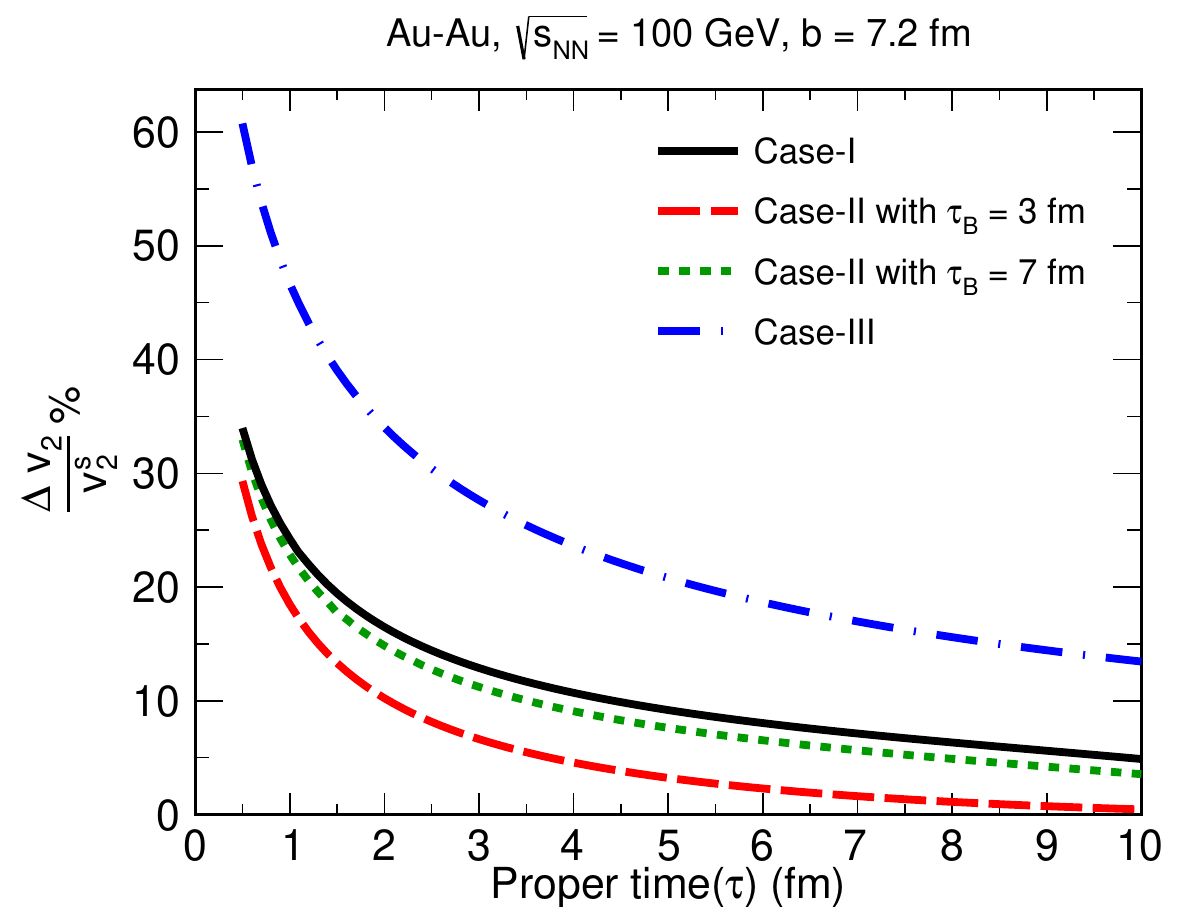}
	\caption{Left: Upper panel represents the elliptic flow coefficient ($v_{2}$) as a function of center of mass energy $\sqrt{s_{NN}}$, the bottom panel represents percentage deviation of elliptic flow coefficient ($v_{2}$) of evolving picture from the static picture as a function of the center of mass energy $\sqrt{s_{NN}}$. Right: Percentage deviation of elliptic flow coefficient ($v_{2}$) of evolving picture from the static picture as a function of proper time ($\tau$). Where, $\Delta v_{2} = v_{2}^{s} - v_{2}^{e}$, the difference of elliptic flow of the evolving case from the static one.}
	\label{Fig-v2}
\end{figure*}

\subsection{Phenomenological Significance}
In the RHICs, azimuthal anisotropy in the momentum space of the produced particles, especially the elliptic flow coefficient $v_2$, is one of the important observable in determining the thermalization of produced medium~\cite{STAR:2003wqp, Bhalerao:2005mm}. 
To estimate the effect of evolution in $v_2$ via thermal conductivity, we use the simple relation~\cite{Bhalerao:2005mm, Drescher:2007cd, Gombeaud:2007ub}
\begin{align}\label{knud}
    v_2=\frac{v_2^{\rm hydro}}{1+\frac{Kn}{Kn^{\rm hydro}}}
\end{align}
where $Kn$ is Knudsen number, $v_2^{\rm hydro}$ is the value of elliptic flow in the hydrodynamic limit, {\i.e.,} at $Kn\rightarrow 0$ limit. The Knudsen number is defined as 
$Kn\equiv \frac{\lambda}{l}$, representing the degree of thermalization of the medium with $\lambda$ as mean free path and $l$ as the characteristic length scale of the system.
One can calculate the mean free path ($\lambda$) using the simple classical relation~\cite{kittel2005introduction}
\begin{equation}\label{Dependence}
\lambda=\frac{3\kappa}{{\rm v}C_V}~,
\end{equation}
which relates the coefficient of thermal conductivity $\kappa$, the specific heat at constant volume $C_V$, and relative speed v.
For the numerical estimation of $v_2$, we have approximated ${\rm v} \approx 1$ and system size $l \approx 3$~fm; $v_2^{\rm hydro} \approx 0.3$ and $Kn^{\rm hydro} \approx 0.7$, which are obtained from the transport calculation~\cite{Gombeaud:2007ub}. 


To obtain $v_2$ as a function of center of mass energy $\sqrt{s_{NN}}$ (in GeV) we use the following parameterization~\cite{Cleymans:2005xv}
\begin{align}\label{par}
    T(\mu_{B}) &= 0.166 - 0.139 \mu_B^2 - 0.053 \mu_B^4, \nn\\
    \mu_{B} &= \frac{1.308}{1 + 0.273\sqrt{s_{NN}}}.
\end{align}
We also parameterize the magnetic field as~\cite{Tuchin2013ie}, 
\begin{align}
    B = \frac{\sqrt{s_{NN}}}{8\pi ~m_N}Ze\frac{b}{R^3} \exp{\left(-\frac{\tau}{\tau_B}\right)}.
\end{align}
Here, $R$ is the radius of colliding ions with electric charge $Ze$, $b$ is the impact parameter, and $m_N = 0.938$~GeV is the nucleon mass. For Au-Au collision at $\sqrt{s_{NN}} = 200$~GeV, with $R=6.38$~fm, $b=7.2$~fm, and $Z = 79$ we get $eB = 9 ~m_\pi^2$ at $\tau = 0$, which is in accordance with the results for RHIC energies~\cite{PhysRevC.85.044907}.

In Fig.~\ref{Fig-v2} (left), we have plotted the elliptic flow coefficient ($v_2$) as a function of the center of mass energy $\sqrt{s_{NN}}$ (in GeV) for Au-Au collision with impact parameter 7.2 fm in the upper panel for all the three cases. We see that for all three cases, the elliptic flow coefficient ($v_2$) decreases as $\sqrt{s_{NN}}$ increases. In Eq.~\ref{par}, we can see that the baryon chemical potential $\mu_{B}$ decreases as $\sqrt{s_{NN}}$ increases. With the increase in values of $\mu_{B}$, the baryon density of medium increases, which alternatively lowers the value of relaxation time $(\tau_R)$ of medium constituents. Therefore, the higher the $\mu_{B}$ value, the lower the thermal conductivity coefficient values ($\kappa$). Hence, it is clear from the Eq.~\ref{knud} that $v_2$ decreases as $\sqrt{s_{NN}}$ increases. The blue solid square represents STAR data of all charged particles for Au-Au collisions with 20\%-30\% centralities, i.e., within the range of impact parameter b = 6.61 - 8.1 fm~\cite{aamodt2010elliptic}, shown for reference purpose only. We can not compare these results with the current estimation, as our results are for the whole medium, including all the charged and uncharged particles.
The bottom panel of the figure shows the percentage deviation of $v_2$ for the evolving picture from the static picture. The positive value of the deviation represents that the elliptic flow in the evolving picture is smaller than the static picture. This deviation is slightly negative at very low energy for all the cases. Later, the deviation value is positive for all the cases throughout the mentioned energy range. So, the temperature evolution reduces the elliptic flow. This effect reduces as we increase the energy. Also, note that the deviation is highly sensitive to the cooling rate, varying from minimum $2\%$ to maximum $24\%$ for different cases. 
In Fig.~\ref{Fig-v2} (right), we have plotted the same as a function of proper time ($\tau$) for Au-Au collision at $\sqrt{s_{NN}} = 100$~GeV and $b=7.2$~fm. Here also, we notice the reduction of $v_2$ in the evolving picture. The effect is very high at the early stage of evolution. In both figures, we see that the effect is comparatively less in the presence of the magnetic field. The slower the decay of the magnetic field, the higher the impact in $v_2$.

Next, we discuss the possibility of incorporating the effect of evolution in different hydrodynamic models. In principle, (3+1)D hydrodynamics models for RHICs show that cooling rates are direction or flow-dependent, for which conductivity will be different in different directions during the evolution. However, in (3+1)D, the complete analytical solution to the hydrodynamics equations is not achievable, and we have to rely on partial analytical solutions or hydrodynamic models where an analytical solution is possible. Here, we use (3+1)D Gubser hydrodynamic~\cite{PhysRevD.82.085027}, which, due to symmetry consideration, reduces to (1+1)D flow.

\subsection{Conductivity in Gubser flow}
Here, we work out thermal conductivity for Gubser hydrodynamics~\cite{PhysRevD.82.085027}. Gubser flow describes a conformally symmetric system that expands cylindrically along the beamline and in the transverse direction (radial flow) in the coordinate system ($\tau, \eta, r, \phi$),
related to Cartesian coordinate system $(t, x, y, z)$ as,
\begin{align}
    &\tau = \sqrt{t^2 - z^2},~~~~\eta = arctanh \frac{z}{t},\nn\\
    &r = \sqrt{x^2 + y^2},~~~~\phi = arctan \frac{y}{x}.
\end{align}
\begin{figure}
	\centering
	\includegraphics[scale=0.42]{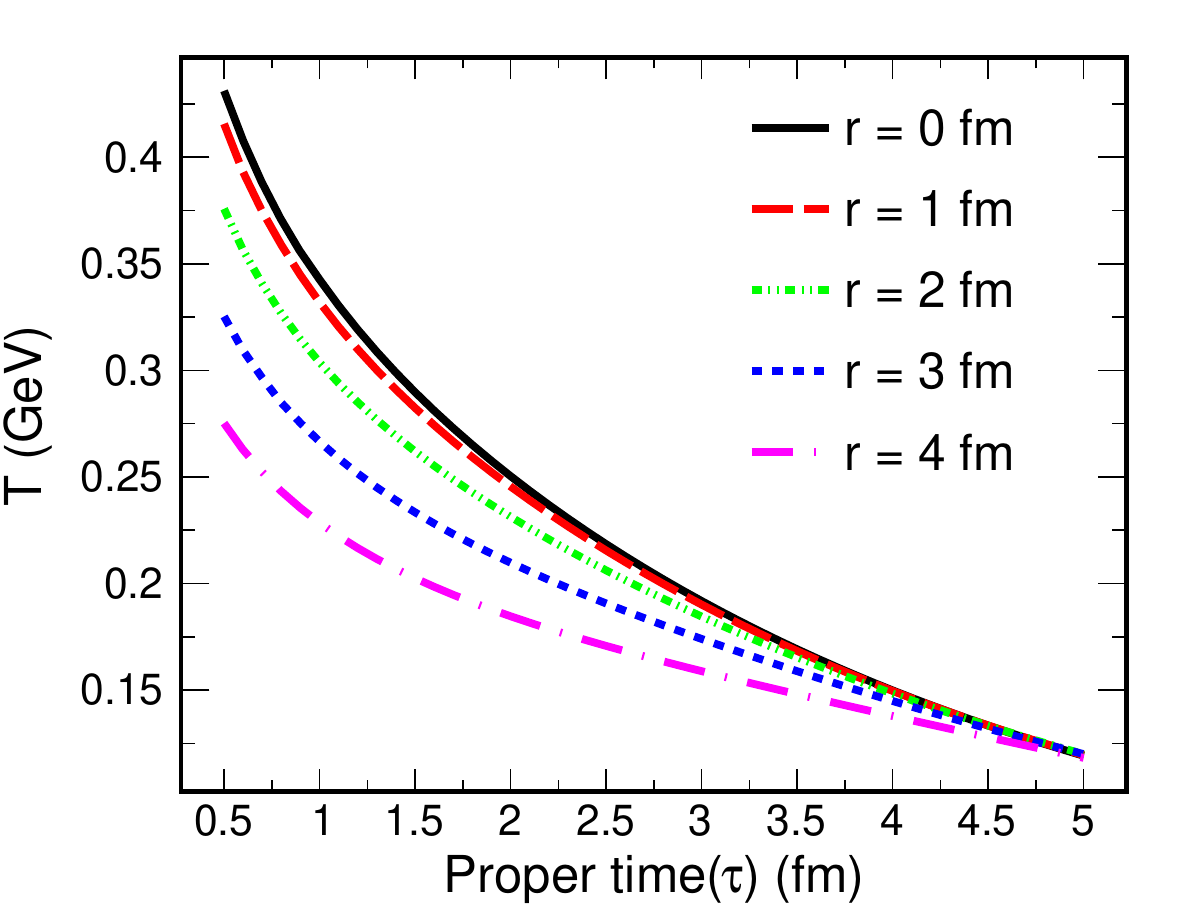}
	\caption{Cooling rate in Gubser flow for different value of $r = 0,1,2,3,4$~fm.}
	\label{Fig-gubt}
\end{figure}
The four-velocity $u^\mu$ is constructed from symmetry consideration with  boost, rotation invariance, and reflection invariance ($\eta \rightarrow - \eta$):
\begin{align}
    &u^{\tau} = \cosh{\kappa} = \gamma_{r},~~~~~~~~~~~~ \frac{u^{r}}{u^{\tau}} = \tanh{\kappa} = v_r,\nn\\
   &u^\eta = u^\phi = 0.
\end{align}
Where $\gamma_r$ = $\frac{1}{\sqrt{1-v_{r}^2}}$ and $v_r$ = $\sqrt{\Vec{v}_{x}^2 + \vec{v}_{y}^2}$, $v_r$ is known as transverse velocity. $\kappa$ is parameterized as $\kappa(\tau, r) = {arctanh} (\frac{2q^{2}\tau r}{1 + q^{2} \tau^{2} +  q^{2} r^{2}})$.
\begin{figure*}
	\centering
	\includegraphics[scale=0.42]{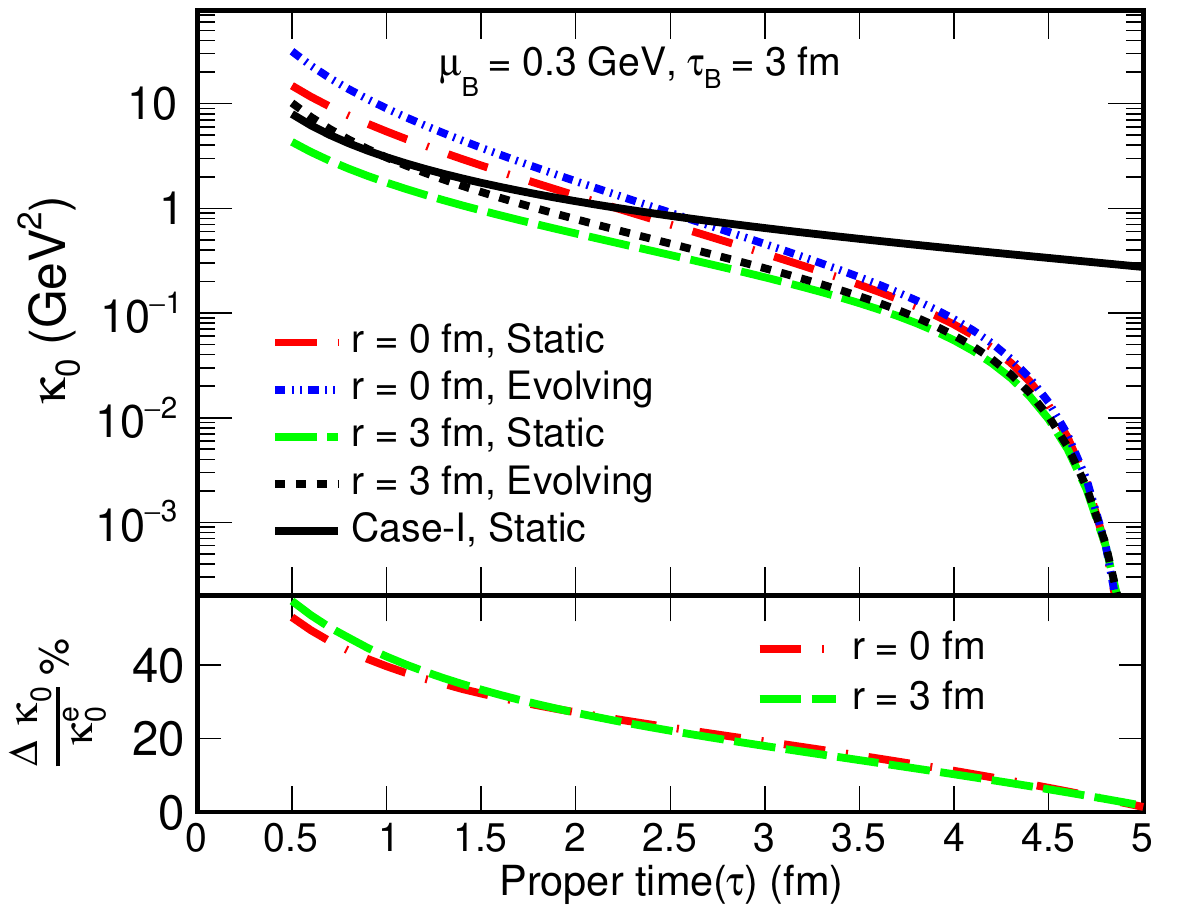}
 \includegraphics[scale=0.42]{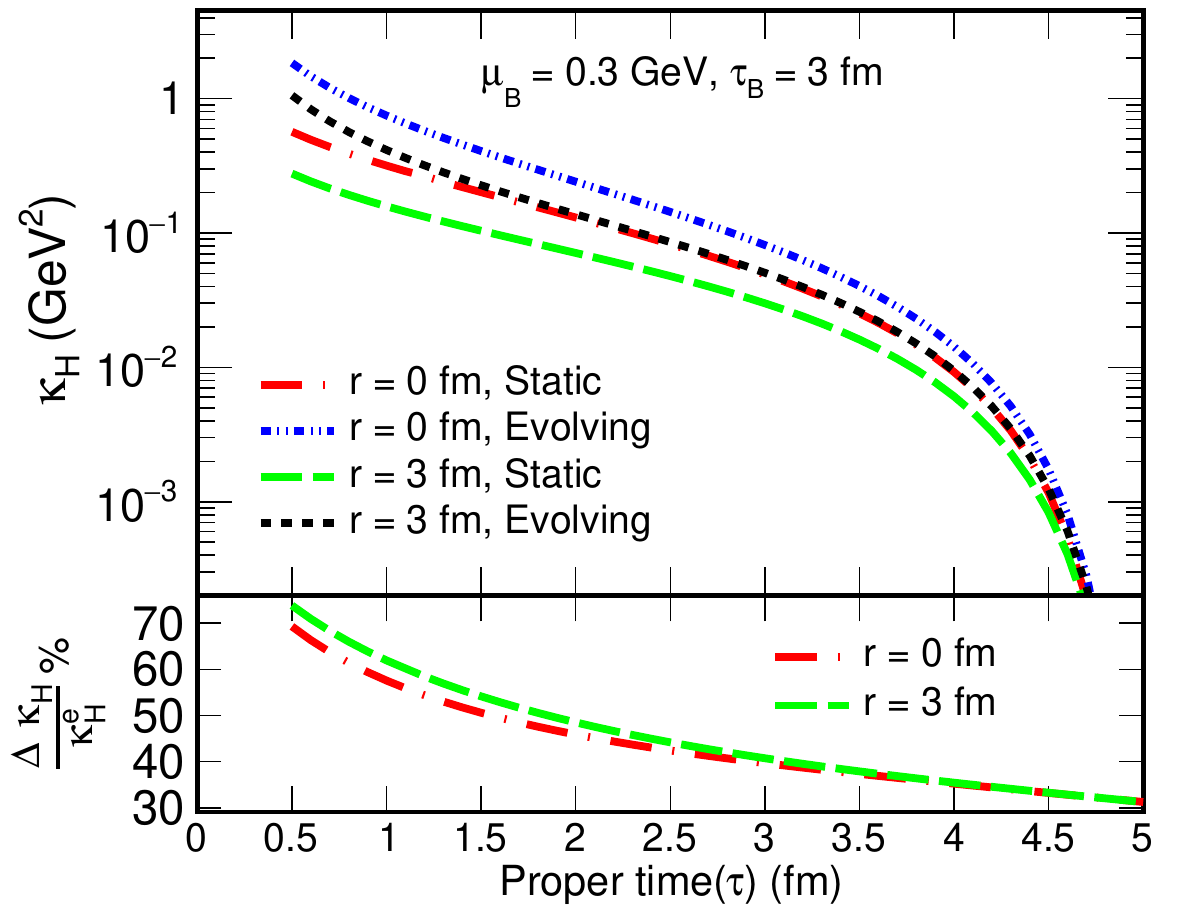}
	\caption{Component of thermal conductivity as a function of proper time ($\tau$). Left: The upper panel represents $\kappa_0$ Component, and the bottom panel represents its percentage deviation (where $\Delta\Tilde\kappa_0 = \kappa_0^{e} - \kappa_0^{s}$) of the static picture from evolving picture; Right: The same as the left figure but for ``Hall-like" Component in Gubser flow.}
	\label{Fig-kag}
\end{figure*}
Now, for the viscous case, with shear viscosity $\tilde\eta = H_0 \epsilon^ {3/4}$, preserving the conformal
symmetry of the system, $p = \epsilon / 3$, temperature can be expressed as\cite{PhysRevD.82.085027},
 \begin{align}\label{gt}
  T &= {1 \over \tau f_*^{1/4}} \Big[ {\frac{{\hat{T}}_0}{(1+g^2)^{1/3}}} + \frac{{\rm H}_0 g}{\sqrt{1+g^2}} \nn\\
  &\left\{1 - (1+g^2)^{1/6} {}_2F_1\left({1 \over 2}, {1 \over 6}; {3 \over 2}; -g^2\right) \right\} \Big]~,  
 \end{align}  
where ${\hat{T}}_0$ is an integration constant, $H_0$ is a dimensionless quantity, $f_*$ is taken to be 11 and ${}_2F_1$ denotes a hypergeometric function. $g$ is the function of radial ($r$) and proper time ($\tau$) coordinates given as,
\begin{align}\label{g}
    g = {1 - q^2 \tau^2 + q^2 r^2 \over 2q\tau}
\end{align}
Here, we use the same parameterization as given in Ref~\cite{PhysRevD.82.085027}, for $1/q = 4.3\,{\rm fm}$ at $\sqrt{s_{\rm NN}} = 200\,{\rm GeV}$ for a central Au-Au collision, parameters turns out to be $\hat{T}_0 = 5.55$ and ${\rm H}_0 = 0.33$. In Fig.~\ref{Fig-gubt}, we have plotted the cooling rate for Gubser flow at different points in the transverse plane \i.e., $r = 0, 1, 2, 3, 4$~fm. 
\begin{figure}
	\centering
	\includegraphics[scale=0.42]{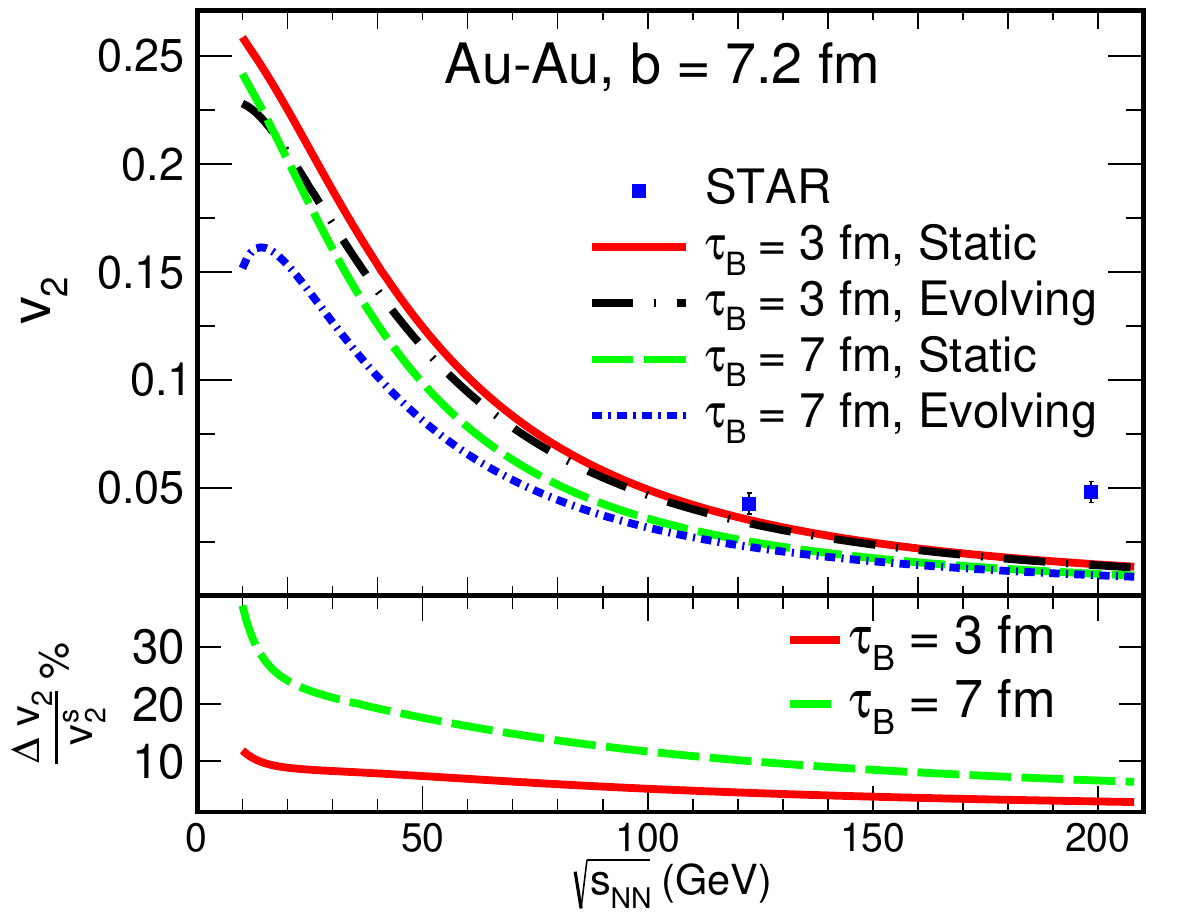}
	 \caption{(Upper panel represents the elliptic flow coefficient ($v_{2}$) as a function of center of mass energy $\sqrt{s_{NN}}$, the bottom panel represents percentage deviation of elliptic flow coefficient ($v_{2}$) of evolving picture from the static picture as a function of the center of mass energy $\sqrt{s_{NN}}$ in Gubser flow.}
	\label{Fig-gubv2}
\end{figure}
Now, we are using this cooling rate to study its effect on the thermal conductivity of the medium. From Eq.~\ref{gt},
\begin{align}
    \frac{\del T}{\del\tau} = &\frac{-T}{\tau} - \zeta (g), \nn\\ 
    \rm{where},\nn\\
    \zeta (g) = &{1 \over \tau f_{*}^{1/4}}\big(\frac{-1}{q} - \frac{g}{\tau}\big)\Big[\frac{2}{3}g(1+g^2)^{-4/3} \hat{T_0}\nn\\
    &- H_0 (1+g^2)^{-1/2}(1-g^{2}(1+g^{2})^{-1}) 
    + H_0 \nn\\ 
    &(1+g^2)^{-1/3}\big\{1-\frac{2g^{2}}{3}(1+g^2)^{-1} {}_2F_1\left({1 \over 2}, {1 \over 6}; {3 \over 2}; -g^2\right) \nn\\
    &-\frac{g^2}{9}{}_2F_1\left({3 \over 2}, {7 \over 6}; {5 \over 2}; -g^2\right)\big\}\Big],\nonumber
\end{align}
therefore,
\begin{align}
    \vec{\nabla} \dot T = &\frac{-1}{\tau}\nabla T + \vec{\nabla} \zeta(g).
\end{align}
Using Eq.~\ref{gt} into Eq.~\ref{II}, we obtained the expressions for components of thermal conductivity, which are
\begin{widetext}
\begin{align}\label{gubkap}
 {\kappa}_0^{e} &= \kappa_0 + \kappa_3, \nn\\
  &= \frac{1}{3T^2} \sum_i g_i \int \frac{d^3|\vk_i|}{(2\pi)^3}\frac{\vec{k}^2_i}{\om_i^2}(\om_i - {\rm b}_i h)^2 \frac{\tau_R^i} {(1+\chi_i + \chi_i^2)}
  \left[1 - \left\{{\tau_R^{i}} \frac{(1+\chi_i)(\chi_i^2-1) + \chi_i^2}{(1+\chi_i)(1+\chi_i^2)}\right\}
\left\{\frac{1}{\tau}\right\}\right] \times f^0_i(1\mp f^0_i),  \nn\\
{\kappa}_H^{e} &= \kappa_1 + \kappa_2 + \kappa_4, \nn\\
&= \frac{1}{3T^2} \sum_i g_i \int \frac{d^3|\vk_i|}{(2\pi)^3}\frac{\vec{k}^2_i}{\om_i^2}(\om_i - {\rm b}_i h)^2 \frac{\tau_R^i \chi_i} {(1+\chi_i + \chi_i^2)}
  \left[1 + \left\{{\tau_R^{i}} \frac{(2+\chi_i)}{(1 + \chi_i)(1+\chi_i^2)}\right\}
\left\{\frac{1}{\tau}\right\}\right]\times f^0_i(1\mp f^0_i).  
\end{align}
\end{widetext} 
 where $\kappa_1=\bar{\kappa}_1 {B}$, $\kappa_2 =\bar{\kappa}_2 \dot{{B}}$, $\kappa_3=\bar{\kappa}_3 \big\{\frac{-1}{\tau} \big\}$, and $\kappa_4=\bar{\kappa}_4\big\{\frac{-1}{\tau} \big\} {B}$,~ with $\chi_i = \frac{\tau_R^i}{\tau_B}$. Here, for the static case, expressions of conductivity are the same as in Eq.~\ref{18}, with temperature profile Eq.~\ref{gt}. Note that expressions of conductivity in Eq.~\ref{gubkap} do not directly depend on $r$ but show the indirect dependence through the Gubser temperature profile, which is a function of $r$ and $\tau$.

In the upper panels of Fig.~\ref{Fig-kag}, we have plotted Ohmic (left) and Hall-like (right) thermal conductivity components using Gubser cooling rate as a function of proper time ($\tau$) for two different values of $r$ = 0 fm, 3 fm. Here, we have plotted the results in the presence of a time-varying magnetic field and the decay parameter $\tau_{B}$ = 3 fm. Here, we see that as $r$ increases, the value of thermal conductivity for both the cases of Ohmic and Hall-like components decreases. The black solid line shown in the upper panel of the left figure represents the thermal conductivity component in the static case for ideal hydrodynamics. In the Gubser case, there is a large deviation in thermal conductivity components from the black solid line. In the Gubser case, the temperature falls faster, leading to this deviation from the Bjorken case in the later evolution time. The bottom panel of both sides shows the deviation of these thermal conductivity components in the static picture from the evolving picture in Gubser flow. A significant deviation, nearly 50\%-60\% for the Ohmic component and 60\%-70\% in the Hall-like component,  is visible in the early stages of the medium at 0.5 fm. However, in the later stages at 5 fm, it is comparatively less nearly 0\%-10\% for the Ohmic component and 20\%-30\% for the Hall-like component.

 In Fig.~\ref{Fig-gubv2}, we study $v_2$ for the Gubser flow case. Here also, we find that $v_2$ reduces in the evolving picture. Though conductivity is $r$ dependent, because of the parametrization in Eq.~(\ref{par}), $v_2$ no longer depends on $r$. It is important to note that the experimental measurements reveal an increasing trend of $v_2$ with $\sqrt{s_{NN}}$~\cite{aamodt2010elliptic}. However, the formula used here to study $v_{2}$ here is not robust.
 This phenomenological study aims to measure a possible effect of the considered evolution picture on $v_2$ due to modification in thermal conductivity instead of absolute $v_2$. A rigorous study with the hydrodynamic simulation is necessary to quantify the exact effect in $v_{2}$ or any other phenomenological quantities due to conductivity modification in the evolving picture.

\section{Summary and Conclusion} 
\label{sec-summary}
In summary, we have calculated the thermal conductivity of an evolving medium where the system temperature changes with time. Heat current is generated when the system has a temperature gradient ($\vec \nabla T$). In the case of QGP created in RHICs, the temperature falls as the medium evolves, which leads to changes in the temperature gradient in the system with time. This change must lead to a finite effect in thermal conductivity and related observables. In this work, we have estimated the effect of $\vec \nabla \dot T$ in thermal conductivity and elliptic flow. QGP created in non-central RHICs faces a substantial magnetic field at an early stage, which decays depending on the electrical properties of the medium. Thermal conductivity can be broken into two independent components in a magnetic field. One is responsible for heat current along the temperature gradient, denoted with $\kappa_0$. Another one is the ``Hall-like" component ($\kappa_H$) along the direction perpendicular to the temperature gradient and magnetic field, which vanishes as soon as the magnetic field diminishes.  
In this work, we have considered a time-varying magnetic field to estimate thermal conductivity. For the decay profile of temperature commonly known as cooling rate, we considered three different hydrodynamic expansion: ideal and first-order hydrodynamics in the absence of magnetic field denoted here by case-I and case-III, respectively, and in the presence of the magnetic field, we have considered the ideal magnetohydrodynamics, denoted by case-II. It is important to note that thermal conductivity is a dissipative quantity and does not appear in ideal hydrodynamics. In principle, causal and stable hydrodynamics, such as second-order MIS theory, should be used to calculate thermal conductivity and cooling rate. However, no analytical solution for cooling rate in (3+1)D hydrodynamics or magnetohydrodynamics exists. Therefore, we have used the ideal and first-order hydrodynamics, where an analytical solution of cooling rate is possible.
Furthermore, in the end, we extend our study to Gubser hydrodynamics. In this case, the temperature is a function of proper time and radial coordinates, affecting thermal conductivity. One can extend the work to study the effect of the evolution in any hydrodynamic model where an analytical or semi-analytical solution is possible.

In this work, we have used a quasiparticle model for QGP, where interaction is taken care of by the thermal mass of the partons. We found that when we take temperature evolution, thermal conductivity increases which is sensitive to the cooling rate. The effect is comparatively higher in the case of first-order cooling rate, where cooling is comparatively slow. In the presence of a magnetic field, the effect of cooling in $\kappa_0$ component is comparatively small and sensitive to the magnetic field decay parameter ($\tau_B$). We have also estimated the effect of temperature evolution on the elliptic flow of QGP via thermal conductivity. We found that elliptic flow reduces due to temperature evolution. The effect is smaller in the presence of a magnetic field, and it increases with increasing the value of the decay parameter. \i.e., slower decay of the magnetic field leads to a higher reduction of elliptic flow due to temperature evolution. 


This is the very first study that includes the effect of temperature evolution in calculating the thermal conductivity of QGP. Despite exciting findings, this work has some limitations or shortcomings that should be resolved for a complete and causally stable result. First, the cooling rate must be calculated using second-order dissipative magnetohydrodynamics. However, in that case, the cooling rate exists between the ideal and first-order hydrodynamics, for which we argue that actual results would not deviate much from our estimations. 
Another point to note here is that the particle's energy gets quantized via the Landau quantization in the presence of a magnetic field, which is not considered here. All these points mentioned above are essential to take care of for hydrodynamical simulation where thermal conductivity is required.

\section*{Acknowledgement}
KS acknowledge the doctoral fellowship from UGC, Government of India. JD and RS gratefully acknowledge the DAE-DST, Govt. of India funding under the mega-science project – “Indian participation in the ALICE experiment at CERN” bearing Project No. SR/MF/PS-02/2021-
IITI (E-37123).

\appendix
\section*{Appendices}
\label{sec-appendix}
\setcounter{equation}{0}
\renewcommand{\thesubsection}{\Alph{subsection}}
\renewcommand{\theequation}{\thesubsection.\arabic{equation}}
\subsection{Appendix: Cooling rate}
\label{appendix1}
The energy-momentum tensor for an ideal magnetohydrodynamics can be
given by ~\cite{HUANG20113075, Giacomazzo:2005jy, Giacomazzo:2007ti}
\begin{align} \label{eq:EMTensor}
{T}^{\mu\nu}=\left(\epsilon +p+{B}^{2}\right){u}^{\mu}{u}^{\nu}-
\left(p+\frac{{B}^{2}}{2}\right){g}^{\mu\nu}-{B}^{\mu}{B}^{\nu},
\end{align}
where $\epsilon, p$, and $u^\mu$ are the fluid energy density, pressure, and four-velocity, respectively. $g^{\mu\nu}$, is the metric tensor of flat spacetime, = $\rm{diag}(1,-1,-1,-1)$. Here $B^{\mu}=\frac{1}{2}
\epsilon^{\mu\nu\alpha\beta}F_{\nu\alpha}u_{\beta}$ is the magnetic field
in the frame moving with the velocity $u_{\beta}$,
$\epsilon^{\mu\nu\alpha\beta}$ is the completely antisymmetric fourth rank Levi-Civita
tensor. The magnetic field
  four-vector $B^{\mu}$ is a spacelike vector with modulus
  $B^{\mu}B_{\mu}=-B^{2}$, and orthogonal to $u^{\mu}$, ie
  $B^{\mu}u_{\mu}=0$, where $B=|\vec{\boldsymbol{B}}|$ and
  $\vec{\boldsymbol{B}}$ is the magnetic field three-vector in the frame
  moving with four-velocity $u^{\mu}$.
The projection of the energy-momentum conservation equation
${\partial}_{\nu}{T}^{\mu\nu}=0$ along the fluid four-velocity,
\begin{align}
{u}_{\mu}{\partial}_{\nu}{T}^{\mu\nu}  =  0,
\end{align}
will express the conservation of energy. 
Using Eq.~\ref{eq:EMTensor}, we get

\begin{align}\label{evol}
\partial_{\tau}\left(\epsilon +  \frac{{B}^{2}}{2}\right) + \frac{\epsilon +  p + {B}^{2}}{\tau} = 0.
\end{align}

Here, we will use Eq.~\ref{evol} to study the temperature evolution of the system for the cases in the absence and presence of a time-varying magnetic field.
\subsubsection{Case-I (B=0)}
In the absence of an external magnetic field Eq.~\ref{evol} reduces into the form,
\begin{align} \label{ideal}
\partial_{\tau}\epsilon + 
\frac{\epsilon +  p}{\tau} &=0.   
\end{align}
Here we consider an ideal equation of state $p=\frac{\epsilon}{3} = aT^{4}$  for QGP. Where 
\begin{align}
a = \left(16 + \frac{21}{2}N_{f}\right)\frac{\pi^{2}}{90}.    
\end{align}
Hence, Eq.~(\ref{ideal}) takes the form,
\begin{align} 
\frac{\partial T}{\partial\tau} &= \frac{-T}{3\tau}. 
\end{align}
The solution of this differential equation with initial conditions of $\tau = \tau_{0}$ at $ T = T_{0}$ is
\begin{align} \label{IdTemp}
T = T_{0} {\left(\frac{\tau_{0}}{\tau}\right)}^{\frac{1}{3}}
\end{align}

\subsubsection{Case-II (B $\not =$ 0)}
In the presence of external time-varying magnetic field Eq.~\ref{evol} simplified into the form,
\begin{align} \label{cb}
\frac{\partial T}{\partial\tau} &= \frac{B(\tau)^2}{12aT^3\tau_B}-\frac{4aT^4+B(\tau)^2}{12aT^3\tau}.
\end{align}
Here $B(\tau) = B_0 \exp(-{\tau / \tau_B})$.
The above equation can be rearranged into the form of a standard Bernoulli's differential equation as
\begin{align}
\frac{\partial T}{\partial\tau} + p(\tau) T = q(\tau) T^{-3},
\end{align}
where p$(\tau)$ = $\frac{1}{3\tau}$, q$(\tau)$ = ($\alpha \beta - \frac{\alpha}{\tau}$)exp(-2$\beta \tau$ ) with $\alpha = \frac{B_0^2}{12a}$ and $\beta = \frac{1}{\tau_B}$.
After solving this differential equation, we get the temperature evolution with time with initial conditions of $\tau = \tau_{0}$ at $ T = T_{0}$ as
\begin{align} \label{Temp}
T &= \Big[T_{0}^4 {\left(\frac{\tau_{0}}{\tau}\right)}^{\frac{4}{3}} + \frac{4\alpha}{{(2\beta\tau)}^{\frac{4}{3}}} \Big\{\Gamma{(4/3, 2\beta\tau)} - \Gamma(4/3, 2\beta\tau_0) \Big\} \nn\\
&~~~~- \frac{2\alpha}{{(2\beta\tau)}^{\frac{7}{3}}}\Big\{\Gamma(7/3, 2\beta\tau) - \Gamma(7/3, 2\beta\tau_0) \Big\}\Big]^\frac{1}{4}.
\end{align}
From Eq.~\ref{cb}, one can easily find,
\begin{align}\label{appcool}
    \vec{\nabla}\dot T =  \left\{\frac{-1}{3\tau}\left(1 - \frac{3B(\tau)^2}{4aT^4} \right)-\frac{B(\tau)^2}{4a\tau_{B}T^4} \right\}\vec{\nabla}T.
\end{align}

\subsection{Appendix: Thermal Conductivity}
\label{appendix2}
\subsubsection{Case-I (B = 0)}
In the case of space-time evolving temperature, the heat current in the rest frame of the fluid can be expressed as
\begin{align}\label{heat}
\vec{I} &= -(\kappa_0 \vec{\nabla}T +\bar{\kappa}_1 \vec{\nabla}\dot T).  \nonumber\\  
\end{align}    
The microscopic definition of heat flow from Eq.~(\ref{1.5}),
\begin{align}\label{mic}
{\vec {I}}_i &= \int \frac{d^3|\vk_i|}{(2\pi)^3} \frac{\vec {k}_i}{\om_i} (\om_i -{\rm b}_i h)\delta f_i.
\end{align}
To find $\delta f$, we solve the BTE in the presence of an external magnetic field under the RTA
\begin{align}\label{5.9A}
	\frac{\del f_i}{\del t} + \frac{\vk_i}{\om_i} \cdot \frac{\del f_i}{\del \vec{x}}
	= -\frac{\delta f_i}{\tau^i_R}~.
\end{align}
We can assume an ansatz of $\delta f_i$ for thermal conductivity as
\begin{align}\label{ans}
\delta f_i=\frac{({\vec{k_i}}.{\vec \Omega}_\kappa )}{T} \frac{\partial f^0_i}{\partial \om_i}. 
\end{align} 
The general form of ${\vec{\Omega}_\kappa}$ is
\begin{align}\label{5.9}
\vec{\Omega}_\kappa =&  \alpha_1\vec{\nabla}T + \alpha_2\vec{\nabla}\dot T 
\end{align} 
The unknown coefficients $\alpha_{i}$ ($i=(1, 2)$) can be obtained by substituting  Eq.~(\ref{5.9}) and Eq.~(\ref{ans}) in Eq.~(\ref{5.9A}),
The first term on left hand side (lhs) of the equation becomes,
\begin{align}\label{5.10}
&  -(\om_i - {\rm b}_{i}\mu)\frac{\dot T}{T}\frac{\partial f^0_i}{\partial \om_i} + \frac{\om_i {\vec{v}_i}}{T}.\Big\{\dot{\alpha_1}{\vec{\nabla}T}+\alpha_1 {{\vec {\nabla}\dot T}}+\dot{\alpha_2} {\vec {\nabla}\dot T}\nn\\
&+\alpha_2 {\vec {\nabla}\ddot T}\Big\}\frac{\partial f^0_i}{\partial \om_i}.
\end{align}
The second term in lhs of the Boltzmann equation leads to, \\
\begin{align}\label{5.11}
\frac{\partial f_i}{\partial x_i} &= \frac{\partial}{\partial x_i}(f^{0}_i +\delta f_i)\nonumber\\ &=\frac{\partial}{\partial x_i}(f^0_i) + \frac{\partial f_i}{\partial x_i}(\delta f_i) \nn\\
&=\frac{\partial}{\partial x_i}\Big(\frac{1}{1\pm\exp{\big(\beta( \om_i - {\rm b}_i \mu_i)\big)}}\Big) + 0\nonumber\\
&= -(\om_i - {\rm b}_ih)\frac{\vec{\nabla}T}{T}\frac{\partial f^0_i}{\partial \om_i}.
\end{align}
Finally, after the substitution of the above results in both sides of Eq.~(\ref{5.9A}), we get
\begin{align}\label{5.10A}
&  -(\om_i - {\rm b}_i\mu)\frac{\dot T}{T}\frac{\partial f^0_i}{\partial \om_i} + \frac{\om_i {\vec{v}_i}}{T}.\Big(\dot{\alpha_1}{\vec{\nabla}T}+\alpha_1 {{\vec {\nabla}\dot T}}+\dot{\alpha_2} {\vec {\nabla}\dot T}\nn\\
&+\alpha_2 {\vec {\nabla}\ddot T}\Big)\frac{\partial f^0_i}{\partial \om_i} -(\om_i - {\rm b}_ih)\frac{\vec{\nabla}T}{T}\frac{\partial f^0_i}{\partial \om_i} = -\frac{\om_i}{T\tau_R^i}\Big(\alpha_1\vec{\nabla}T\nn\\ 
&+ \alpha_2\vec{\nabla}\dot T\Big)\frac{\partial f^0_i}{\partial \om_i}.
\end{align}
In the current analysis, we consider only the terms with first-order derivatives of the fields and neglect higher-order derivative terms. The comparison of the coefficients of ${\vec {v}}\cdot{\vec {\nabla} T}$ and ${\vec {v}}\cdot{\vec {\nabla} \dot T}$ on both sides of the above equation leads to,
\begin{align}\label{alphas}
 \dot{\alpha_1} &=  -\bigg(\frac{1}{\tau_R^i}\alpha_1 -\frac{\om_i - {\rm b}_ih}{\om_i}\bigg), \nn\\
 \alpha_2 &= -\tau_R^i \alpha_1. 
\end{align}
Eq.~(\ref{alphas}) is a first-order differential equation. After solving it we get,
\begin{align}
 {\alpha_1} &=  -{\tau_R^i}\bigg(\frac{\om_i - {\rm b}_i h}{\om_i} -  \exp(-\tau/\tau_R)\bigg),   
\end{align}
After substituting Eq.~(\ref{ans}) Eq.~(\ref{5.9}) into Eq.~(\ref{mic}) we get,
\begin{align} \label{cur}
{\vec {I}}_i &= \frac{1}{3T}\sum_i \int \frac{d^3|\vk_i|}{(2\pi)^3\om_{i}} {\vec {k}}_i^2 (\om_i -{\rm b}_ih)\{\alpha_1\vec{\nabla}T 
+ \alpha_2\vec{\nabla}\dot T\}\nn\\
&\hspace{60mm}(\frac{\partial f_i^0}{\partial \om_i}).
\end{align}
Comparison of Eq.~(\ref{cur}) and Eq.~(\ref{heat}) leads to,
\begin{align} \label{kapa}
\kappa_0 &=  -\frac{1}{3T} \sum_{\rm baryons} g_i \int \frac{d^3|\vk_i|}{(2\pi)^3\om_{i}} k^2_i(\om_i - {\rm b}_ih)^2 \alpha_1(\frac{\partial f_i^0}{\partial \om_i}),\nn\\ 
\bar\kappa_1 &=  -\frac{1}{3T} \sum_{\rm baryons} g_i \int \frac{d^3|\vk_i|}{(2\pi)^3\om_{i}} k^2_i(\om_i - {\rm b}_ih)^2 \alpha_2(\frac{\partial f_i^0}{\partial \om_i}).
\end{align}
\subsubsection{Case-II (B $\not=$ 0)}
In the presence of a time-varying magnetic field heat current in the fluid rest frame can be expressed as~\cite{PhysRevD.106.034008}
\begin{align}\label{IIap}
\vec{I} &= -\big\{\kappa_0 \vec{\nabla}T +\bar{\kappa}_1 (\vec{\nabla}T \times \vec{B}) +\bar{\kappa}_2 (\vec{\nabla}T \times \dot{\vec{B}}) + \bar{\kappa}_3  \vec{\nabla}\dot T \nonumber\\
 &~~~+\bar{\kappa}_4 (\vec{\nabla}\dot T \times\vec{B}) \big\}.~  
\end{align} 
We can express the three vector form of heat current in terms of microscopic quantities as 
\begin{align}\label{1.5aa}
{\vec {I}}_i = \int \frac{d^3|\vk_i|}{(2\pi)^3} \frac{\vec {k}_i}{\om_i} (\om_i -{\rm b}_i h)\delta f_i.
\end{align}
To find $\delta f_i$, we solve the BTE in the presence of an external magnetic field under the RTA
\begin{align}\label{BTE-RTA-ThA}
	\frac{\del f_i}{\del \tau} + \frac{\vk_i}{\om_i} \cdot \frac{\del f_i}{\del \vec{x}} + q_i \left(\frac{\vk_i}{\om_i} \times \vec{B}\right) \cdot \frac{\del f_i}{\del \vec{k_i}}
	= -\frac{\delta f_i}{\tau^i_R}~.
\end{align}
where a general form of ${\vec{\Omega}_\kappa}$ up to first order time derivative of $\vec B$ can be expressed as
\begin{align}\label{1.7a}
\vec{\Omega}_\kappa &= \alpha_1\vec{B}+ \alpha_2\vec{\nabla}T+ \alpha_3(\vec{\nabla}T \times \vec{B})+\alpha_4 \dot{\vec{B}} +\alpha_5(\vec{\nabla}T \times \dot{\vec{B}})\nn\\
&+\alpha_6\vec{\nabla}\dot T+\alpha_7(\vec{\nabla}\dot T \times \vec{B})+\alpha_8(\vec{\nabla}\dot T \times \dot{\vec{B}}).
\end{align} 
The unknown coefficients $\alpha_{i}$ ($i=(1, 2.....8)$) can be obtained by substituting  Eq.~(\ref{1.7a}) and Eq.~(\ref{ans})  in Eq.~(\ref{BTE-RTA-ThA}).

The first term in left-hand side (lhs) of the Eq.~(\ref{BTE-RTA-ThA}) becomes,
\begin{align}\label{5.10}
&\frac{\om_i {\vec{v}_i}}{T}.\Big\{\dot{\alpha_1 }{\vec {B}}+\alpha_1 \dot{{\vec {B}}}+\dot{\alpha_2} {\vec {\nabla}T} + {\alpha_2} {\vec {\nabla}\dot T}+\alpha_3 ({\vec {\nabla}T}\times \dot{{\vec {B}}})\nonumber\\
&+\dot{\alpha_3 }({\vec {\nabla}T} \times {\vec {B}})+ {\alpha_3 }({\vec {\nabla} \dot T} \times {\vec {B}})
+\dot{\alpha_4} \dot{\vec{B}}+\alpha_4 \ddot{\vec{B}}\nonumber\\
&+\dot{\alpha_5 }({\vec {\nabla}T} \times \dot{{\vec {B}}})
+\alpha_5 (\vec{\nabla}\dot T \times{\vec{B}}) +\alpha_5 (\vec{\nabla}T \times\ddot{\vec{B}})+ \dot {\alpha_6} {\vec {\nabla}\dot T}\nonumber\\ 
&+ {\alpha_6} {\vec {\nabla}\ddot T} 
+ \dot {\alpha_7} ({\vec {\nabla}\dot T}\times{{\vec {B}}}) +{\alpha_7 }({\vec {\nabla}\ddot T} \times {\vec {B}})+ {\alpha_7 }({\vec {\nabla} \dot T} \times \ddot {\vec {B}}) \nonumber\\
&+ \alpha_8 ({\vec {\nabla}\ddot T}\times \dot{{\vec {B}}}) +\dot{\alpha_8 }({\vec {\nabla}\dot T} \times \dot {\vec {B}})+ {\alpha_8 }({\vec {\nabla}T} \times \ddot {\vec {B}})\nn\\\
&-(\om_i - {\rm b}_i\mu)\frac{\dot T}{T}\Big\}\frac{\partial f^0_i}{\partial \om_i}.
\end{align}
The second term in lhs of the Boltzmann equation leads to, 
\begin{align}\label{5.11}
\frac{\partial f_i}{\partial x_i} &= \frac{\partial}{\partial x_i}(f^{0}_i +\delta f_i)\nonumber\\ &=\frac{\partial}{\partial x_i}(f^0_i) + \frac{\partial f_i}{\partial x_i}(\delta f_i) \nn\\
&=\frac{\partial}{\partial x_i}\Big\{\frac{1}{1\pm\exp{\big(\beta( \om_i - {\rm b}_i \mu_i)\big)}}\Big\} + 0\nonumber\\
&= -(\om_i - {\rm b}_ih)\frac{\vec{\nabla}T}{T}\frac{\partial f^0_i}{\partial \om_i}.
\end{align}
The third term in lhs leads to,
\begin{align}\label{5.12}
\frac{\partial f_i}{\partial k_i} &= \frac{\partial f^0_i}{\partial k_i} + \frac{\partial \delta f_i}{\partial k_i}\nonumber\\ 
&= \vec{v_i}\frac{\partial f^0_i}{\partial \om_i} + 
\frac{\vec {\Omega}_\kappa}{T}  \frac{\partial f^0_i}{\partial \om_i}.
\end{align}
The identity $({\vec {v_i}} \times {\vec {B}}) \cdot \vec{v_i}\frac{\partial f^0_i}{\partial \om_i} = 0$. Thus we are left with only $\frac{\vec {\Omega}_\kappa}{T}  \frac{\partial f^0_i}{\partial \om_i}$ term of the above equation. Hence,
\begin{align}\label{5.13}
 \frac{\partial f_i}{\partial k_i} &= \frac{1}{T}\big\{-\alpha_2 q_{i}\vec{v_i}\cdot(\vec{\nabla}T \times \vec{B}) +\alpha_3 q_{i}\vec{v_i}\cdot\vec{\nabla}T(\vec{B}\cdot\vec{B})-\alpha_3 q_{i}\nonumber\\
 &\vec{v_i}\cdot\vec{B}
 (\vec{B}\cdot\vec{\nabla}T)
+\alpha_5q_{i}\vec{v_i}\cdot\vec{\nabla}T(\dot{\vec{B}}\cdot\vec{B}) -\alpha_5 q_{i}\vec{v_i}\cdot\dot{\vec{B}}\nn\\
&(\vec{B}\cdot\vec{\nabla}T)
 -\alpha_6 q_{i}\vec{v_i}\cdot(\vec{\nabla}\dot T \times \vec{B}) +\alpha_7 q_{i}\vec{v_i}\cdot\vec{\nabla}\dot T (\vec{B}\cdot\vec{B})\nonumber\\
 &-\alpha_7 q_{i}
 \vec{v_i}\cdot\vec{B}
 (\vec{B}\cdot\vec{\nabla}\dot T)+\alpha_8q_{i}\vec{v_i}\cdot\vec{\nabla}\dot T(\dot{\vec{B}}\cdot\vec{B}) -\alpha_8 q_{i}\nn\\
 &\vec{v_i}\cdot\dot{\vec{B}}
(\vec{B}\cdot\vec{\nabla}\dot T)\big\}\frac{\partial f^0_i}{\partial \om_i}. 
 \end{align}
Finally, after the substitution of the above results on both the sides of Eq.~(\ref{BTE-RTA-ThA}), 
\begin{widetext}
the left side becomes
\begin{align}\label{5.14}
&=\frac{\om_i {\vec{v}_i}}{T}.\Big\{\dot{\alpha_1 }{\vec {B}}+\alpha_1 \dot{{\vec {B}}}+\dot{\alpha_2} {\vec {\nabla}T} + {\alpha_2} {\vec {\nabla}\dot T}+\alpha_3 ({\vec {\nabla}T}\times \dot{{\vec {B}}})
+\dot{\alpha_3 }({\vec {\nabla}T} \times {\vec {B}})+ {\alpha_3 }({\vec {\nabla} \dot T} \times {\vec {B}})
+\dot{\alpha_4} \dot{\vec{B}}+\alpha_4 \ddot{\vec{B}}
+\dot{\alpha_5 }({\vec {\nabla}T} \times \dot{{\vec {B}}})\nn\\
&+\alpha_5 (\vec{\nabla}\dot T \times{\vec{B}}) +\alpha_5 (\vec{\nabla}T \times\ddot{\vec{B}})+ \dot {\alpha_6} {\vec {\nabla}\dot T} 
+ {\alpha_6} {\vec {\nabla}\ddot T} 
+ \dot {\alpha_7} ({\vec {\nabla}\dot T}\times{{\vec {B}}}) +{\alpha_7 }({\vec {\nabla}\ddot T} \times {\vec {B}})+ {\alpha_7 }({\vec {\nabla} \dot T} \times \ddot {\vec {B}}) 
+ \alpha_8 ({\vec {\nabla}\ddot T}\times \dot{{\vec {B}}}) \nn\\
&+\dot{\alpha_8 }({\vec {\nabla}\dot T} \times \dot {\vec {B}})+ {\alpha_8 }({\vec {\nabla}T} \times \ddot {\vec {B}})
-(\om_i - {\rm b}_i\mu)\frac{\dot T}{T}\Big\}
-(\om_i -{\rm b}_i h)\vec{v_i}\cdot\frac{\vec{\nabla}T}{T}
 +\frac{1}{T}\big\{-\alpha_2 q_{i}\vec{v_i}\cdot(\vec{\nabla}T \times \vec{B}) +\alpha_3 q_{i}\vec{v_i}\cdot\vec{\nabla}T(\vec{B}\cdot\vec{B})\nn\\
 &-\alpha_3 q_{i}
 \vec{v_i}\cdot\vec{B}
 (\vec{B}\cdot\vec{\nabla}T)
+\alpha_5q_{i}\vec{v_i}\cdot\vec{\nabla}T(\dot{\vec{B}}\cdot\vec{B}) -\alpha_5 q_{i}\vec{v_i}\cdot\dot{\vec{B}}
(\vec{B}\cdot\vec{\nabla}T)
 -\alpha_6 q_{i}\vec{v_i}\cdot(\vec{\nabla}\dot T \times \vec{B}) +\alpha_7 q_{i}\vec{v_i}\cdot\vec{\nabla}\dot T (\vec{B}\cdot\vec{B})\nn\\
 &-\alpha_7 q_{i}
 \vec{v_i}\cdot\vec{B}
 (\vec{B}\cdot\vec{\nabla}\dot T)+\alpha_8q_{i}\vec{v_i}\cdot\vec{\nabla}\dot T(\dot{\vec{B}}\cdot\vec{B}) -\alpha_8 q_{i}
 \vec{v_i}\cdot\dot{\vec{B}}
(\vec{B}\cdot\vec{\nabla}\dot T)\Big\},
\end{align}   
and the right side becomes
\begin{align}
= -\frac{\om_i}{T\tau_R^i}\Big\{\alpha_1\vec{B}+ \alpha_2\vec{\nabla}T+ \alpha_3(\vec{\nabla}T \times \vec{B})
+\alpha_4 \dot{\vec{B}}
+\alpha_5(\vec{\nabla}T \times \dot{\vec{B}})
+\alpha_6\vec{\nabla}\dot T+\alpha_7(\vec{\nabla}\dot T \times \vec{B})
+\alpha_8(\vec{\nabla}\dot T \times \dot{\vec{B}})\Big\}\frac{\partial f^0_i}{\partial \om_i}.    
\end{align}
\end{widetext}
In the current analysis, we consider only the terms with first-order derivatives of the fields and neglect higher-order derivative terms. The comparison of the coefficients of ${\vec {v}}\cdot{\vec {B}}$, ${\vec {v_i}}\cdot{\vec {\nabla}T}$, ${\vec {v_i}}\cdot({\vec {\nabla}T \times \dot{\vec{B}}})$,  ${\vec {v_i}\cdot \dot{\vec{B}}}$, ${\vec {v_i}}\cdot({\vec {\nabla}T\times \vec{B}})$, ${\vec {v_i}}\cdot{\vec {\nabla} \dot T}$, and ${\vec {v_i}}\cdot({\vec {\nabla}\dot T\times \vec{B}})$ on both sides, gives us $\dot{\alpha_1}$, $\alpha_4$, $\dot{\alpha_2}$, $\alpha_5$, $\dot{\alpha_3}$, $\alpha_6$, $\alpha_7$respectively as 
\begin{align}\label{dl1}
  \dot{\alpha_1} &= -\frac{1}{\tau_R^i}\alpha_1+\frac{q_{i}\big\{(\vec{B}\cdot\vec{\nabla}T)\alpha_3 + (\vec{B}\cdot\vec{\nabla}\dot T)\alpha_7\big\}}{\om_i}, \nn\\
  \dot{\alpha_2} &=  -\bigg\{\frac{1}{\tau_R^i}\alpha_2 +(\frac{q_{i}(\vec{B}\cdot\vec{B}-\tau_R^i\vec{B}\cdot\dot{\vec{B}})}{\om_i})\alpha_3 -\frac{\om_i - {\rm b}_ih}{\om_i}\bigg\},\nn \\
   \dot{\alpha_3} &= -\frac{1}{\tau_R^i}\alpha_3 +\frac{q_{i}}{\om_i}\alpha_2,\nn \\
  {\alpha_4} &=-{\tau_R^i}\Big\{\alpha_1+\frac{\tau_R^i q_{i} (\vec{B}\cdot\vec{\nabla}T)}{\om_i} \alpha_5 \Big\},\nn\\    
  \alpha_5 &= -\tau_R^i \alpha_3,\nn\\
  \alpha_6 &= -\tau_R^i \alpha_2 - \frac{\tau_R^i q_{i} \vec{B}^2}{\om_i} \alpha_7 ,\nn\\  
  \alpha_7 &= -\tau_R^i \alpha_3 + \frac{\tau_R^i q_{i}}{\om_i} \alpha_6.
\end{align}
Here, $\dot{\alpha_1}$, $\dot{\alpha_2}$, $\dot{\alpha_3}$ from Eq.~(\ref{dl1}) can be expressed in terms of matrix equation as
\begin{equation}\label{5.16}
 \frac{d X}{d t}= AX +G,   
\end{equation}
where the matrices take the following forms
\begin{align}
 X=\begin{pmatrix}
\alpha_1\\\alpha_2\\\alpha_3
\end{pmatrix},\nn
\end{align}
\begin{align}
A=\begin{pmatrix}
-\frac{1}{\tau_R^i} &- \frac{\tau_R^{i}\frac{q_i}{\om_i}Q}{P} &\frac{q_{i}}{\om_i}(\vec{B}\cdot\vec{\nabla}T)- \frac{Q}{P}\\
 0 &-\frac{1}{\tau_R^i} & -\frac{q_{i} F^2}{\om_i}\\
0 &\frac{q_{i}}{\om_i} &  -\frac{1}{\tau_R^i}, 
\end{pmatrix}
\end{align}
\begin{align}
G =\begin{pmatrix}
0\\ \frac{\om_i - {\rm b}_ih}{\om_i}\\ 0
\end{pmatrix},\nonumber
\end{align}
with $F = \sqrt{B(B-\tau_R^i \dot{B})}$, $P = {1 + \tau_R^{i2} \frac{q_{i}^2\vec{B}^2}{\om_i^2}}$, $Q = \frac{\tau_R^{i}q_i}{\om_i}(\vec{B}\cdot\vec{\nabla}\dot T) $. Eq.~(\ref{5.16}) can be solved by diagonalizing the matrix $A$ and using the method of the variation of constants. The eigenvalues corresponding to matrix $A$ are,\\
$\lambda _ j $ = $-\frac{1}{\tau_R^i} + a_j i\frac{q_{i} F}{\om_i}$,~~
with $a_1=0$, $a_2=-1$, $a_3=1$.\\
Hence, one can write the linear independent solutions corresponding to the homogeneous part of differential  Eq.~ (\ref{5.16}) in terms of its eigenvectors as
\begin{align}
y_1=\begin{pmatrix}
 e^{\eta_1}\\ 0 \\ 0
\end{pmatrix},
y_2=\begin{pmatrix}
\zeta e^{\eta_2} \\-iF e^{\eta_2}\\ e^{\eta_2}
\end{pmatrix},
y_3=\begin{pmatrix}
-\zeta e^{\eta_3} \\iF e^{\eta_3}\\ e^{\eta_3}
\end{pmatrix}.\nonumber
\end{align}
Where,
\begin{align}\label{5.19}
\zeta &= \frac{i(\vec{B}\cdot\vec{\nabla}T)}{F}+ \frac{Q}{P}\left(\tau_R^i + \frac{i\om_i}{q_i F}\right),\nn\\
\eta_j &= -\frac{\tau}{\tau_R^i} +a_j\frac{q_{i} i}{\om}\int F d\tau.
\end{align}
Therefore, the fundamental matrix for Eq.~(\ref{5.16}) is 
\begin{align}\label{5.20}
&Y=\begin{pmatrix}
e^{\eta_1}& \zeta e^{\eta_2} & -\zeta e^{\eta_3}\\
0 & -iF e^{\eta_2} & iF e^{\eta_3}\\
0 & e^{\eta_2} & e^{\eta_3}\\
\end{pmatrix},
\end{align}
We seek a particular solution of the equation with a given form of Eq.~(\ref{5.16})
is
\begin{align}\label{5.22}
Y_p &= YU,
\end{align}
where $U$ is a column matrix of unknowns. 
\begin{align}\label{5.23}
U=\begin{pmatrix}
u1\\ u2\\ .\\ .\\ .\\ u_n\\
\end{pmatrix}.
\end{align}
From Eq.~(\ref{5.22}) and Eq.~(\ref{5.16}) we can see that $Y_p$ is a column matrix with the same coefficients as that of matrix X. Further, the differentiation of Eq.~(\ref{5.22}) with respect to time gives us
$Y_p^{'}$ = $Y^{'}U + YU^{'}$, where $Y^{'} = AY$. 
Hence, 
$Y_p^{'}$ = $AY_p + YU^{'}$.
Comparison of above equation with Eq.~(\ref{5.16}) tells us 
\begin{align}\label{5.24}
 G = YU^{'}.   
\end{align}
The determinant of matrix $Y$ is $-2iFe^{\eta}$. Then, 
\begin{align}\label{5.25}
u^{'}_1 &= \frac{1}{-2iFe^{\eta}}
det
\begin{pmatrix}
0& \zeta e^{\eta_2} & -\zeta e^{\eta_3}\\
\frac{\om_i - {\rm b}_ih}{\om} & -iF e^{\eta_2} & iF e^{\eta_3}\\
0 & e^{\eta_2} & e^{\eta_3}\\
\end{pmatrix}, \nonumber\\
&= \frac{e^{-\eta_1}}{iF} (\frac{\om_i - {\rm b}_ih}{\om_i}). \nonumber
\\u^{'}_2 &= \frac{1}{-2iFe^{\eta}} 
 det
\begin{pmatrix}
e^{\eta_1}& 0 & -\zeta e^{\eta_3}\\
0 & \frac{\om_i - {\rm b}_ih}{\om_i} & iF e^{\eta_3}\\
0 & 0 & e^{\eta_3}\\
\end{pmatrix},\nonumber\\
&= \frac{-e^{-\eta_2}}{iF} (\frac{\om_i - {\rm b}_ih}{2\om_i}).\nonumber
\\u^{'}_3 &= \frac{1}{-2iFe^{\eta}}
det
\begin{pmatrix}
e^{\eta_1}& \zeta e^{\eta_2} & 0\\
0 & -iF e^{\eta_2} & \frac{\om_i - {\rm b}_ih}{\om_i}\\
0 & e^{\eta_2} & 0\\
\end{pmatrix},\nonumber\\
&= \frac{e^{-\eta_3}}{iF} (\frac{\om_i -{\rm b}_ih}{2\om_i}). 
\end{align}
After integrating $u^{'}_1$, $u^{'}_2$ and $u^{'}_3$ with respect to time , we get the matrix $U$ as\\
\begin{align}\label{5.26}
U &=
\begin{pmatrix}
-i \frac{\om_i - {\rm b}_ih}{\om_i} \xi_1\\
i \frac{\om_i - {\rm b}_ih}{2\om_i} \xi_2\\
-i \frac{\om_i - {\rm b}_ih}{2\om_i} \xi_3\\
 \end{pmatrix}.
 \end{align}
Where $\xi_j = \int \frac{e^{-\eta_{j}}}{F} d\tau$. \\
\\After substituting the above value of $U$ in Eq.~(\ref{5.22}) one obtain,

\begin{align}
Y_p &= 
\begin{pmatrix}
e^{\eta_1}& \zeta e^{\eta_2} & -\zeta e^{\eta_3}\\
0 & -iF e^{\eta_2} & iF e^{\eta_3}\\
0 & e^{\eta_2} & e^{\eta_3}\\
\end{pmatrix}
\begin{pmatrix}
-i \frac{\om_i - {\rm b}_ih}{\om_i} \xi_1\\
i \frac{\om_i - {\rm b}_ih}{2\om_i} \xi_2\\
-i \frac{\om_i - {\rm b}_ih}{2\om_i} \xi_3\\
\end{pmatrix}, \nonumber\\
\begin{pmatrix}
\alpha_1\\\alpha_2\\\alpha_3
\end{pmatrix}
 &= 
\begin{pmatrix}
e^{\eta_1}& \zeta e^{\eta_2} & -\zeta e^{\eta_3}\\
0 & -iF e^{\eta_2} & iF e^{\eta_3}\\
0 & e^{\eta_2} & e^{\eta_3}\\
\end{pmatrix} 
\begin{pmatrix}
c_1\\
c_2\\
c_3\\
\end{pmatrix}.\nonumber
\end{align}
Hence,\\
\begin{align}\label{talpha3}
\alpha_1 &= c_1e^{\eta_1} + c_2 \zeta e^{\eta_2} - c_3 \zeta e^{\eta_3}, \nonumber\\
\alpha_2 &= -c_2iFe^{\eta_2} + c_3iFe^{\eta_3},\nonumber\\
\alpha_3 &= c_2e^{\eta_2} + c_3e^{\eta_3}.   
\end{align}
The functions $c_1(\tau)$, $c_2(\tau)$ and $c_3(\tau)$ can be defined as $c_1 =-i\frac{(\om_i - {\rm b}_ih)}{\om_i}  \xi_1$,  $c_2 =i\frac{(\om_i - {\rm b}_ih)}{2\om_i} \xi_2$ and $c_3 =-i\frac{(\om_i - {\rm b}_ih)}{2\om_i} \xi_3$.
 For time-varying field we get\\
$ F = B\sqrt{1+\frac{\tau_R^i}{\tau_B}}$ , ~~$\int F d\tau$ = $B\tau\sqrt{1+\frac{\tau_R^i}{\tau_B}}$.\\
Hence, it leads to the given form, 
\begin{align}\label{5.29}
    \eta_j &= -\frac{\tau}{\tau_R^i} +a_ji\frac{\sqrt{1+\frac{\tau_R^i}{\tau_B}}}{\tau_B}~\tau,\nn \\ 
\xi_j &= \frac{1}{\sqrt{1+\frac{\tau_R^i}{\tau_B}} B_0} \frac{e^{\Big(\frac{1}{\tau_R^i} +\frac{1}{\tau_B}-a_ji\frac{\sqrt{1+\frac{\tau_R^i}{\tau_B}}}{\tau_B} \Big)\tau}}{\Big( \frac{1}{\tau_R^i} +\frac{1}{\tau_B}-a_ji\frac{\sqrt{1+\frac{\tau_R^i}{\tau_B}}}{\tau_B} \Big)}.
\end{align}
Using Eq.~(\ref{1.5aa}) we get,
\begin{align}
{\vec {I}}_i &= \frac{1}{3T}\sum_i \int \frac{d^3|\vk_i|}{(2\pi)^3\om_{i}} {\vec {k}}_i^2 (\om_i -{\rm b}_ih)\{\alpha_1\vec{B}+ \alpha_2\vec{\nabla}T\nn\\
 &+\alpha_3(\vec{\nabla}T\times \vec{B})
 +\alpha_4 \dot{\vec{B}} +\alpha_5(\vec{\nabla}T\times \dot{\vec{B}})+\alpha_6\vec{\nabla}\dot T\nn\\
 &+\alpha_7(\vec{\nabla}\dot T \times \vec{B})+\alpha_8(\vec{\nabla}\dot T \times \dot{\vec{B}})\}(\frac{\partial f_i^0}{\partial \om_i}).\label{5.32}
\end{align}\\
where,
\begin{align}
\alpha_2 &= \frac{(\om_i - {\rm b}_i h)}{\om_i} \tau_R^i ~\frac{1}{(1+\chi_i + \chi_i^2)},  \nn \\ 
\alpha_3 &= \frac{(\om_i - {\rm b}_ih)}{\om_i B } \tau_R^i ~\frac{\chi_i}{(1+\chi_i)(1+\chi_i + \chi_i^2)}.\nn\\ 
\alpha_6 &= \frac{(\om_i - {\rm b}_ih)}{\om_i}  ~\frac{\tau_R^{i2}}{(1+\chi_i)(1+\chi_i + \chi_i^2)}\nn\\
&\hspace{22mm}\times\big\{\frac{\left((1+\chi_i)(\chi_i^2-1)\right) + \chi_i^2}{1+\chi_i^2}\big\}\nn\\
\alpha_7 &= -\frac{(\om_i - {\rm b}_ih)}{\om_i B } \tau_R^{i2} ~\frac{\chi_i + \chi_i(1+\chi_i)}{(1+\chi_i)(1+\chi_i + \chi_i^2) (1+ \chi_i^2)}.\label{alphathree}
\end{align}
Where $\chi_i = \frac{\tau_R^i}{\tau_B}$.\nn\\
Compare the coefficients of Eq.~(\ref{IIap}) and Eq.~(\ref{5.32})
\begin{align}
\kappa_0 &=  -\frac{1}{3T} \sum_{\rm baryons} g_i \int \frac{d^3|\vk_i|}{(2\pi)^3\om_{i}} k^2_i(\om_i - {\rm b}_ih)^2 \alpha_2(\frac{\partial f_i^0}{\partial \om_i}),\nn\\ 
\bar\kappa_1 &=  -\frac{1}{3T} \sum_{\rm baryons} g_i \int \frac{d^3|\vk_i|}{(2\pi)^3\om_{i}} k^2_i(\om_i - {\rm b}_ih)^2 \alpha_3(\frac{\partial f_i^0}{\partial \om_i}),\nn \\
\bar\kappa_2 &=  -\frac{1}{3T} \sum_{\rm baryons} g_i \int \frac{d^3|\vk_i|}{(2\pi)^3\om_{i}} k^2_i(\om_i - {\rm b}_ih)^2 (-\tau_R^i \alpha_3) (\frac{\partial f_i^0}{\partial \om_i}),\nn\\
\bar\kappa_3 &=  -\frac{1}{3T} \sum_{\rm baryons} g_i \int \frac{d^3|\vk_i|}{(2\pi)^3\om_{i}} k^2_i(\om_i - {\rm b}_ih)^2 \alpha_6(\frac{\partial f_i^0}{\partial \om_i}),\nn\\
\bar\kappa_4 &=  -\frac{1}{3T} \sum_{\rm baryons} g_i \int \frac{d^3|\vk_i|}{(2\pi)^3\om_{i}} k^2_i(\om_i - {\rm b}_ih)^2 \alpha_7(\frac{\partial f_i^0}{\partial \om_i}).\label{5.33C}
\end{align}

\bibliographystyle{apsrev4-2}
\bibliography{reference}
\end{document}